\documentclass{aastex}

\usepackage{emulateapj5}
\usepackage{graphics}

\newcommand{\mc}[1]{\multicolumn{2}{c}{#1}}
\newcommand{\mct}[1]{\multicolumn{3}{c}{#1}}
\newcommand{\tna}{\tablenotemark{a}}

\newcommand{\tnb}{\tablenotemark{b}}

\newcommand{\rtwo}{{r_{200}}}

\newcommand{\mr}[1]{\mathrm{#1}}

\newcommand{\m}{$^{-1}$}

\newcommand{\hsq}{h_{100}^{-2}}

\newcommand{\ngroup}{$41$}

\newcommand{\blskip}{}
\newcommand{\mb}[1]{\mathbf{#1}}
\newcommand{\grad}{\mb{\nabla}}
\newcommand{\dd}[2]{\frac{d {#1}}{d {#2}}}
\newcommand{\dlndln}[2]{\frac{d \, {\ln{#1}}}{d \, {\ln{#2}}}}

\newcommand{\ene}{\mathcal{E}}
\newcommand{\cross}{\mb{\times}}
\newcommand{\grouprange}{1.8--2.2}
\newcommand{\clusterrange}{1.6--2.0}
\newcommand{\rlim}{R_\mr{lim}}

\begin{document}
\blskip

\submitted{Preprint date: \today}

\title{A Redshift Survey of Nearby Galaxy Groups: The Shape of the
Mass Density Profile}

\submitted{Accepted for publication in the Astrophysical Journal}

\author{Andisheh Mahdavi\altaffilmark{1}} \affil{Institute for
 Astronomy, University of Hawaii, 2680 Woodlawn Drive, Honolulu, HI 96822}

\and

\author{Margaret J.  Geller\altaffilmark{2}} \affil{Smithsonian
Astrophysical Observatory, 60 Garden St., Cambridge, MA 02138}

\altaffiltext{1}{amahdavi@ifa.hawaii.edu}
\altaffiltext{2}{mgeller@cfa.harvard.edu}

\shorttitle{Shape of the Matter Density}
\shortauthors{Mahdavi \& Geller}

\begin{abstract}
We constrain the mass profile and orbital structure of nearby groups
and clusters of galaxies. Our method yields the joint probability
distribution of the density slope $n$, the velocity anisotropy
$\beta$, and the turnover radius $r_0$ for these systems. The
measurement technique does not use results from N-body simulations as
priors. We incorporate 2419 new redshifts (included here) in the
fields of \ngroup\ systems of galaxies with $z < 0.04$. The new groups
have median velocity dispersion $\sigma=360$ km s\m. We also use 851
archived redshifts in the fields of 8 nearly relaxed clusters with $z
< 0.1$. Within $R \lesssim 2 \rtwo$, the data are consistent with a
single power law matter density distribution with slope $n =
$\grouprange\ for systems with $\sigma < 470$ km s\m\ and $n =
$\clusterrange\ for those with $\sigma > 470$ (95\% confidence). We
show that a simple, scale-free phase space distribution function
$f(E,L^2) \propto (-E)^{\alpha-1/2} L^{-2 \beta}$ is consistent with
the data as long as the matter density has a cusp. Using this DF,
matter density profiles with constant density cores ($n=0$) are ruled
out with better than 99.7\% confidence.

\end{abstract}

\section{Introduction}

The galaxies in a cluster account for less than 10\% of its total
mass. X-ray observations of the hot intracluster gas support the
notion that most of the matter is distributed in a smooth, dark
halo. It is of clear interest to use observations to deduce not only
the total mass, but also the structure of this halo. For example, the
logarithmic derivative of the density can be compared with N-body
simulations of structure formation, allowing constraints on dark
matter physics. These simulations show that the inner regions of halos
composed of collisionless dark matter have power law density profiles
$\rho \propto r^{-n}$ with $n \sim 1-1.5$
\citep{NFW,Fukushige97,Nakano99,Moore99}. However, if the dark matter
particles interact through the weak or the strong force, constant
density cores develop within a fraction of a Hubble time
\citep{Burkert00,Dave01, Balberg02}. Measurements of the slope $n$ in
groups and clusters of galaxies are thus a potential a test of the
collisionless nature of dark matter.

The methods for determining the masses of clusters fall into three
broad categories: fluid dynamics, lensing, and stellar dynamics. The
fluid dynamical methods, which rely on the X-ray properties of the
intracluster medium, presuppose that the gas is either in hydrostatic
equilibrium or in a time-independent cooling flow. Because measuring
the gas temperature involves complex modeling of X-ray spectra, these
methods offer at best $\sim 50$ resolution elements across the
brightest X-ray sources. Furthermore, recent X-ray observations
indicate that the central regions of cooling flow clusters cannot be
described self-consistently by the standard models, leaving the shape
of the mass profile within $\sim 200$ kpc of the cluster center
uncertain\footnote{Throughout this paper we use $H_0 = 100$ km
s\m\ Mpc\m.}  \citep{Markevitch99,Soker01,Fabian01}.

Lensing models detect either strong or weak gravitational deflection
of the light from distant sources. Strong lensing allows direct
estimation of the surface mass density from the spectacular but rare
giant arcs, e.g. CL0024+16 \citep{Tyson98}. Weak lensing estimates
depend on statistical reconstruction of the surface density from
distortions in the shapes of field galaxies
\citep{Clowe00,Sheldon01}. These techniques are attractive because
they make no assumptions about the equilibrium state of the
cluster. However, they could suffer contamination from the large scale
structure surrounding the cluster \citep*{Metzler01}. This
contamination may account for the disagreement between X-ray and
lensing masses, and makes the determination of the true shape of the
density profile difficult.

Stellar dynamical methods grow out of the century-old tradition,
beginning with Jeans and Eddington, of modeling the structure of star
clusters. Here, instead of addressing a self-gravitating system of
stars, one regards galaxies as point masses adrift in a larger sea of
dark matter. The fundamental dynamical problem is then to calculate
the spherically symmetric gravitational potential that causes the
observed motions of the galaxies. Obviously, the analogy to stellar
systems is far from perfect; the ratio of the size of a galaxy to that
of its host cluster is typically $\approx$ 15 kpc / 1.5 Mpc $ = 0.01$,
whereas for stars in a galaxy it is $\approx 10^{-12}$. As a result,
interaction cross sections are much larger in galaxy
clusters. Furthermore, although stellar systems contain as many as
$10^{12}$ members, clusters rarely have more than $\approx 500$
luminous members, and the most abundant systems---groups of
galaxies---have closer to $\approx 30$ \citep{Carlberg96, Mahdavi99}.

Still, if correctly applied, stellar dynamics can trace the
gravitational potential of clusters at the largest and smallest scales
as well as the other available techniques. Modeling of spherical
infall patterns \citep{Geller99,Rines00} can map the mass profile
outside $\approx 5$ Mpc. The equilibrium models we consider are
sensitive to the shape of the dark matter halo in the innermost
regions of clusters and groups.

The most popular equilibrium techniques make use of the moments of the
data to constrain the depth or shape of the cluster potential. For
example, the virial theorem yields an estimate of the total mass of
the cluster \citep{Heisler85,Biviano93,Oegerle95,Carlberg96,Girardi00}:
\begin{eqnarray}
M & = & \frac{3 \pi}{2 G}\sum_{i=1}^N v_{z,i} ^2 R_h, \label{eq:virial} \\
R_h & = & \sum_{i=1}^N R_i^{-1} \\
v_{z,i} & = & \frac{c z_i - \sum_{i=1}^N c z_i/N}
{1+\sum_{i=1}^N  z_i} \label{eq:vzi},
\end{eqnarray}
where $v_z$ is the line-of-sight velocity in the center of mass frame
\citep{Danese}, $z_i$ is the redshift of the $i$th galaxy, $R_i$ is
its projected distance from the cluster center, $c$ is the speed of
light, and $G$ is Newton's constant.

The velocity moments may also be used to infer the structure of the
potential in greater detail, using either analytic galaxy distribution
functions \citep{KentGunn} or the Jeans equation for a collisionless
stellar system \citep{Fabricant89,BinneyTremaine87,Hartog96,Carlberg97},
\begin{equation}
\frac{1}{\nu} \dd{(\nu \sigma_r^2)}{r} + \frac{2 
(\sigma_r^2-\sigma_t^2)}{r}
= - \dd{\Phi}{r},
\label{eq:jeanstwo}
\end{equation}
where $r$ is the true three-dimensional distance from the cluster
center, $\nu(r)$ is the number distribution of the galaxies,
$\sigma_t(r)$ and $\sigma_r(r)$ are the tangential and the radial
velocity dispersions, and $\Phi$ is the gravitational potential.  To
interpret the data, one usually chooses a form for $\Phi(r)$ and the
anisotropy parameter $\beta(r) \equiv 1-\sigma_t^2/\sigma_r^2$, and
projects the solution to the Jeans equation to obtain $\sigma_z(R)$,
the theoretical line-of-sight velocity dispersion profile. This
profile may be compared with a real cluster by splitting the galaxies
into radial bins and calculating $\sigma_z^2 \propto \sum v_z^2$ in each
bin.

There are several disadvantages to using the velocity moments to
constrain $\Phi$. First, the quality of the observed velocity
dispersion profile $\sigma_z(R)$ is usually poor for clusters. Even
with several hundred velocities, dividing the galaxies into radial
bins produces a noisy profile that is not very informative about the
radial variation of $\sigma_z$. Second, even in the ideal limit of a
perfectly observed $\sigma_z(R)$, vastly different combinations of
$\Phi(r)$ and $\beta$ can yield similar solutions to equation
(\ref{eq:jeanstwo}) \citep{BinneyTremaine87}. Third, even if a unique
solution is possible (e.g., by assuming a constant mass-to-galaxy
ratio, $\grad^2 \Phi \propto \nu$), there is no guarantee that the
solutions satisfy the requirement that the phase space density of the
member galaxies be everywhere positive or zero \citep{van00}. Finally,
it is desirable to avoid binning the velocities altogether in order to
make as powerful a use of the scarce data as possible.
We therefore turn to maximum likelihood methods, which avoid some of
the problems of the Jeans equations as applied to discrete systems
\citep{Merritt93,van00}.  Constructing and maximizing a suitable
likelihood function guarantees a positive definite galaxy phase space
distribution.

Here we apply the maximum likelihood method to nearby ($z \lesssim
0.1$) systems of galaxies. Our goal is to derive joint constraints on
$\Phi$ and $\beta$ for poor groups of galaxies as well as for rich
clusters. To this end we conduct deep optical observations of the
RASSCALS X-ray emitting galaxy groups \citep{Mahdavi00}. We also
assemble a catalog of published redshifts in 8 nearby relaxed clusters
of galaxies (\S \ref{sec:data}). Using these samples, we construct an
ensemble group and ensemble cluster which serve to constrain the
galaxy phase space distribution in each type of system (\S
\ref{sec:phase}). This distribution then serves as a maximum
likelihood estimator of the gravitational potential and of the orbital
structure of the galaxy population. In particular, we calculate joint
five-dimensional confidence volumes in the central anisotropy, central
matter density slope, total mass, matter core radius, and interloper
fraction (\S \ref{sec:dynamics}). We discuss the implications of our
work (\S\ref{sec:discuss}) and conclude (\S \ref{sec:conclude}).

\section{Data}
\label{sec:data}

Here we describe the data acquisition. First, we discuss our
observations of primarily poor groups of galaxies. From these data we
obtain a statistically complete sample of 2419 optical spectra
in the fields of \ngroup\ systems. We also assemble a redshift catalog
of galaxies in relaxed rich clusters from the literature.

\subsection{New Observations}

Our redshift measurement targets (Table \ref{tbl:groups}) are drawn
from \cite{Mahdavi00}, who cross-correlate the Center for Astrophysics
Redshift Survey with the ROSAT All-Sky Survey (RASS) to construct the
RASSCALS catalog. The RASSCALS are a statistically complete,
redshift-selected sample of galaxy groups with at least 5 members
brighter than a Zwicky blue magnitude $m_Z = 15.5$, comparable to a
red magnitude $m_R
\approx 14.4$. The groups lie within a redshift range $3000 \le c z
\le 12000$ km s\m, and have either X-ray luminosities or upper limits
derived from the RASS. The diffuse X-ray emission coincident with the
detected groups suggests that they are physically bound systems of
galaxies.

For this study, we select all X-ray emitting RASSCALS with right
ascension $8 < \alpha_{2000} < 16$, or $\alpha_{2000} > 22$ and
$\alpha_{2000} < 2$, and declination $\delta_{2000} > -8$.  We
restrict ourselves to systems well away from the local supercluster,
$c z > 5100$ km s\m, and attempt to ensure that they are group-like by
limiting ourselves to X-ray luminosities $10^{42} < L_X < 10^{43}
\hsq$ erg s\m. We also omit the \cite{Hickson82} compact groups 51 and
58. In addition to this 93\% complete sample of 30 X-ray emitting
groups, we randomly select 11 groups without RASS X-ray emission, all
located within the same ($\alpha_{2000},\delta_{2000},z$)
bounds. There are \ngroup\ systems in all.

To constrain the variation of the velocity dispersion, and hence the
mass, with radius, it is necessary to identify group members to a
large distance from the center of the gravitational potential.  We
therefore identify galaxies brighter than $m_R \approx 15.4$ within
1.5 Mpc of the center of all \ngroup\ groups as spectroscopic
targets. We include all of these galaxies in our spectroscopic
survey. Occasionally, we do not reobserve galaxies with a known
redshift $z$ such that $| c z - c \bar{z}| > 4000$ km s\m, where
$\bar{z}$ is the redshift of the group as listed in
\cite{Mahdavi00}. In other words, we sometimes ignore
spectroscopically confirmed foreground and background galaxies.

The total area of the sky covered by our spectroscopic survey is
$\approx 184$ square degrees. Thus the photometric calibration of
plate-derived magnitudes is difficult, and the magnitude limit is only
an estimate from the Palomar Observatory Sky Survey (POSS)
photographic plates, with a typical scatter of $0.5$ mag from system
to system. Because we view the galaxies as test particles embedded in
a gravitational potential, completeness is not as important as it
would be if we wanted to measure the group luminosity function.
Nevertheless, our sampling uses a strict magnitude cutoff in the field
of each group, with no biases against galaxies close together on the
sky.

We obtained spectra of the 2419 target galaxies with the FAST
single object spectrograph \citep{Fabricant98} on the 1.5m Tillinghast
Reflector at Whipple Observatory on Mount Hopkins, AZ.  Using a 300
line mm\m\ grating, a 3$\arcsec$ slit, and a spectral coverage of
3940\AA\ centered at 5500\AA, we had a typical resolution of 4\AA. A
data reduction pipeline incorporating the cross-correlation of the
spectra with rest-frame galaxy templates
\citep{TonryDavis79,Kurtz92,Kurtz98} gave redshifts with an average
uncertainty of $\pm 40$ km s\m.  Roughly 20\% of these data were
reported in \cite{Mahdavi99}; for completeness, we list all 2419
redshifts in Table \ref{tbl:gals}.

\begin{figure*}
\begin{center}
\resizebox{7in}{!}{\includegraphics{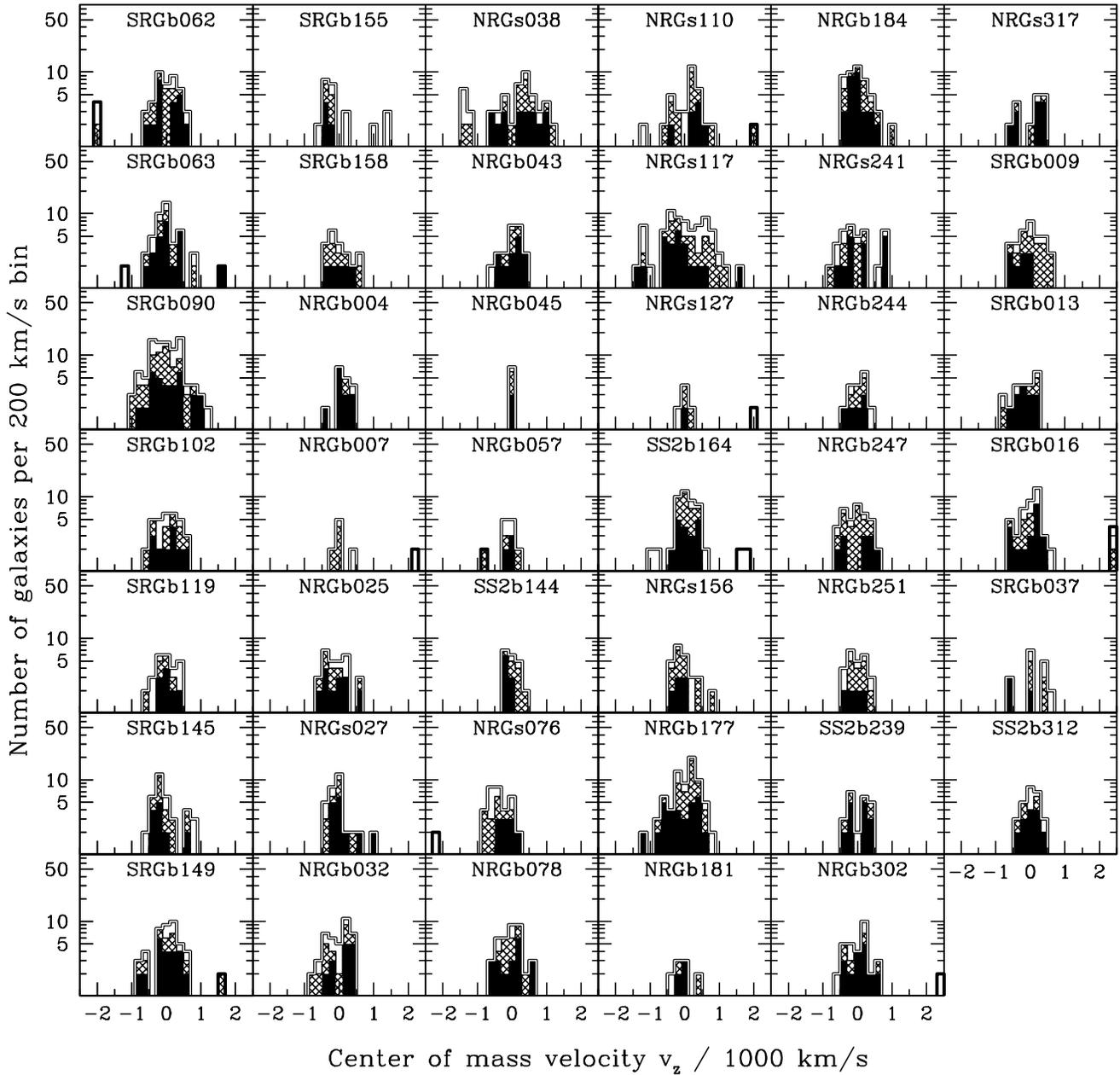}}
\figcaption{Velocity histograms of the galaxies observed in the \ngroup\
groups.  Heavily outlined histograms indicate interlopers identified
by the membership algorithm (\S\protect\ref{sec:membership}). Filled,
hatched, and empty histograms represent member galaxies within 0.5,
1.0, and 1.5 Mpc of the group center. 
\label{fig:hist}}
\end{center}
\end{figure*}

\subsection{Clusters}
\label{sec:clusters}

To compare the mass distribution of poor systems of galaxies and
nearby rich clusters, we assemble a catalog of clusters close to
dynamical equilibrium. This requirement is important, because the
techniques we use below (and in fact all other methods relying on
equilibrium dynamics) assume that the phase space distribution of the
member galaxies is constant with time. Of course, at some level all
clusters have residual substructure; our best hope is to examine
those systems with the smallest departures from equilibrium.

It is surprising just how rare equilibrium systems are. For example,
of the 10 brightest objects in the ROSAT Bright Cluster Sample
\citep[BCS]{Ebeling98}, only two clusters, Abell 1795 and Abell 2029,
make it to our list in Table \ref{tbl:clusterstwo}. Our criteria are
not stringent: we require that published studies of the X-ray emitting
gas and the cluster galaxy population lack clear evidence of ongoing
mergers. For example a strong bimodality in the X-ray or optical light
distribution is a signal that the system is not close to dynamical
equilibrium. We also require a large number ($>70$) of measured
velocities in the field of each cluster; we search the NASA
Extragalactic Database (NED) for published redshifts within
1.5 Mpc of the cluster center.

The dynamical state of some of the clusters in our list is
ambiguous. For example, the central dominant galaxy in Abell 2029 has
a large peculiar velocity $\sim 500$ km s\m\ in the center of mass
frame \citep{Oegerle95}, but the cluster has a regular X-ray and
optical morphology and shows no statistically significant substructure
\citep{Slezak94,Oegerle95}. Many of the clusters we select contain
cooling flows, which should arise only under equilibrium conditions
\citep{Allen98b}. However, not all cooling flow clusters have
equilibrium galaxy populations; the most obvious example is Abell 85,
which exhibits quite an irregular galaxy velocity structure
\citep{Durret98}. Abell 2199, another cooling flow cluster, is regular
in both the X-ray and optical data within 1.5 Mpc. However, it is on
course to collide with Abell 2197 and several other small galaxy
groups located beyond 2 Mpc \citep{Rines01}.

The clusters in our catalog have a mean redshift $0.04$, and therefore
complement the clusters of the Canadian Network for Observational
Cosmology \citep[CNOC]{Yee96,Carlberg96,Carlberg97,van00}, which are
at $z \approx 0.3$ on the average. The two samples differ in
character, however: at least six of the CNOC clusters exhibit obvious
substructure or asymmetry in X-ray images \citep{Neumann97,Lewis99},
and possibly also in weak lensing maps \citep{Clowe00}. The CNOC mass
determinations rely on the assumption that equilibrium techniques
average over the substructure without biasing the final result. To a
lesser extent, this assumption underlies our study as well; the
absence of substructure still does not guarantee that the galaxy phase
space distribution has arrived at a steady state.

\section{The Phase Space Diagram}
\label{sec:phase}

In this section we isolate the group and cluster members from
foreground and background galaxies. To conduct the dynamical modeling,
we combine the systems into two ensembles.

\subsection{Membership}
\label{sec:membership}

A redshift survey in the field of a particular cluster is always
contaminated by foreground and background galaxies. Some outliers in
velocity space are easily identifiable. Unrelated galaxies with
velocities close to the cluster mean, however, are not as cleanly
separated. Nor is it possible to identify a redshift range $(z_1,z_2)$
which contains only the cluster members, because
interlopers---unrelated galaxies with large peculiar motions---could
well be projected into this redshift range.

We adopt a twofold approach to the membership problem. Here we exclude
the most probable foreground and background galaxies by the method of
\cite*{Zabludoff90}. Then in \S\ref{sec:interlopers} we follow
\cite{van00} in folding the existence of the remaining interlopers
directly into our dynamical model.

The \cite{Zabludoff90} method consists entirely in making sure that in
a sorted list of member velocities $v_{z,i}$, no two neighbors have
$\Delta v_z > \sigma_z$, where $\sigma_z$ is the line-of-sight
velocity dispersion of the cluster. This approach is not the most
sophisticated method available. There is an extensive literature on
the use of nonparametric adaptive kernels to reconstruct the cluster
velocity distribution \citep{Pisani93}.  However, the interpretation
of the smooth probability distributions produced by the adaptive
kernels is not straightforward. In particular, the frequent occurrence
of slightly double-peaked probability distributions makes it difficult
to tell exactly which galaxies should be considered cluster
members. The \cite{Zabludoff90} method is also nonparametric given a
redshift range, and it reports the cluster membership unambiguously.

To implement this method, we begin with the known mean system velocity
$c \bar{z}$, from \cite{Mahdavi00} for groups, and from NED for the
clusters. We remove all galaxies in our survey with $|c z - c \bar{z}|
> 3000$ km s\m\ for each cluster. We then calculate $\bar{z}$ and
$\sigma_z$, given by
\begin{equation}
\sigma^2_z = \frac{1}{N-1}\sum_{i = 1}^N v_{z,i}^2,
\label{eq:sigmaz}
\end{equation}
where $v_{z,i}$, the center-of-mass velocity of each galaxy, is shown
in equation (\ref{eq:vzi}). We sort the $v_{z,i}$, and remove all
velocity-space neighbors with a velocity difference $|\Delta v_z| >
\sigma_z$. We repeat the procedure until no galaxy is rejected. Tables
\ref{tbl:groups} and \ref{tbl:clusterstwo} list the final membership
counts. This list almost certainly contains additional interlopers,
which we treat probabilistically in \S \ref{sec:dynamics}. A total of
1428 of the 2419 surveyed galaxies (60\%) are group members. Of the
979 additional redshift taken from the literature, 851 (86\%) are
cluster members.

\subsection{The Ensemble Systems}
\label{sec:reject}

The data for any given system are scarce; there are on the average
$\sim 35$ members per system in the group database. It is therefore
not possible to measure the shape of the matter density for individual
systems. Our approach is to combine the groups and clusters into two
ensembles. If the groups and clusters are self-similar---i.e., they
are differently scaled versions of the same archetypal system---the
aggregate yields the maximum information about the shape of the matter
distribution. If on the other hand the systems in each ensemble are
not self-similar, the ensemble itself represents an average over the
various matter distributions, and the derived results ought to reflect
the range of shapes present in each ensemble.

How are the systems to be combined? Ideally, both the velocity and the
position of galaxies in each system ought to be scaled to a common
value. We could choose a scaling such that the total virial mass of
each system as given by equation (\ref{eq:virial}) is the same:
$\tilde{v} \equiv v_z/\sigma_z$, and $\tilde{R} \equiv
R/R_h$. However, interlopers make the radial scaling problematic. In
practice, scaling by $R_h$ introduces noise in the data because of its
sensitivity to interlopers and substructure \citep{Mahdavi99}. A more
suitable scaling radius is $\rtwo$, the radius at which the cluster
density is 200 times the critical density of the universe.

A common method for calculating $\rtwo$, due to \cite{Carlberg97}, is
to assume that the mass profile $M(r) \propto r$ in the vicinity of
$\rtwo$. Although observed X-ray and lensing cluster mass profiles are
not proportional to $r$ everywhere \citep{Markevitch99,Clowe00}, the
linear relation does appear to be reasonable near $\rtwo$. This
assumption, together with the virial theorem (with $R_h$ replaced by
$\rtwo$), yields the following adopted scalings:
\begin{eqnarray}
\tilde{v}_z & \equiv & v_z/\sigma_z \\
\rtwo & = & \frac{\sqrt{3} \sigma_z}{10 H_0(z)}  \\
\tilde{R} & \equiv &  R/\rtwo
\label{eq:r200}
\end{eqnarray}
where $H_0(z)$ is the Hubble constant at the redshift of the cluster.

Using the velocity and radius scalings, we combine the groups and the
clusters into two different ensembles---one ``high $\sigma$'' and one
``low $\sigma$'' ensemble. We do not combine all the data into one
ensemble because lower mass systems of galaxies are thought to exhibit
``similarity breaking,'' or departures from the scaling laws that
pertain to rich clusters of galaxies. For example \cite{Ponman99},
\cite{Helsdon00}, \cite{Mahdavi00}, and \cite{Mahdavi01} show that for
low mass systems of galaxies, the relationships between the X-ray
luminosity $L_X$, X-ray temperature $T$, and velocity dispersion
$\sigma$ differs from that of rich clusters. It would be interesting
to see if corresponding differences in the matter distributions of
these systems exist. Some high resolution N-body simulation have
reported differences in the shape of the matter distribution for low-
and high-mass systems \citep{Jing00}.

Another reason to break the data into two ensembles is
completeness. In order for the analysis to be valid, the systems
assigned to an ensemble need to have redshift surveys complete to the
same scaled radius. Because of the fixed physical limits of our survey
(1.5 Mpc), low-$\sigma$ (less massive) groups are complete to a much
larger value of $\rtwo$ than are high-$\sigma$ (more massive)
systems. To make efficient use of the data, it is best to have two
different ensembles with different completeness limits. For this
purpose it is useful to define $N(\rlim)$ as the total number of
galaxies within $\rlim$ in all systems that are complete to
$\rlim$. By maximizing $N(\rlim)$ as a function of $\rlim$ for each of
the two ensembles, we ensure that we are making the best possible use
of the data while adhering to the completeness requirement.

\begin{figure*}
\begin{center}
\resizebox{7in}{!}{\includegraphics{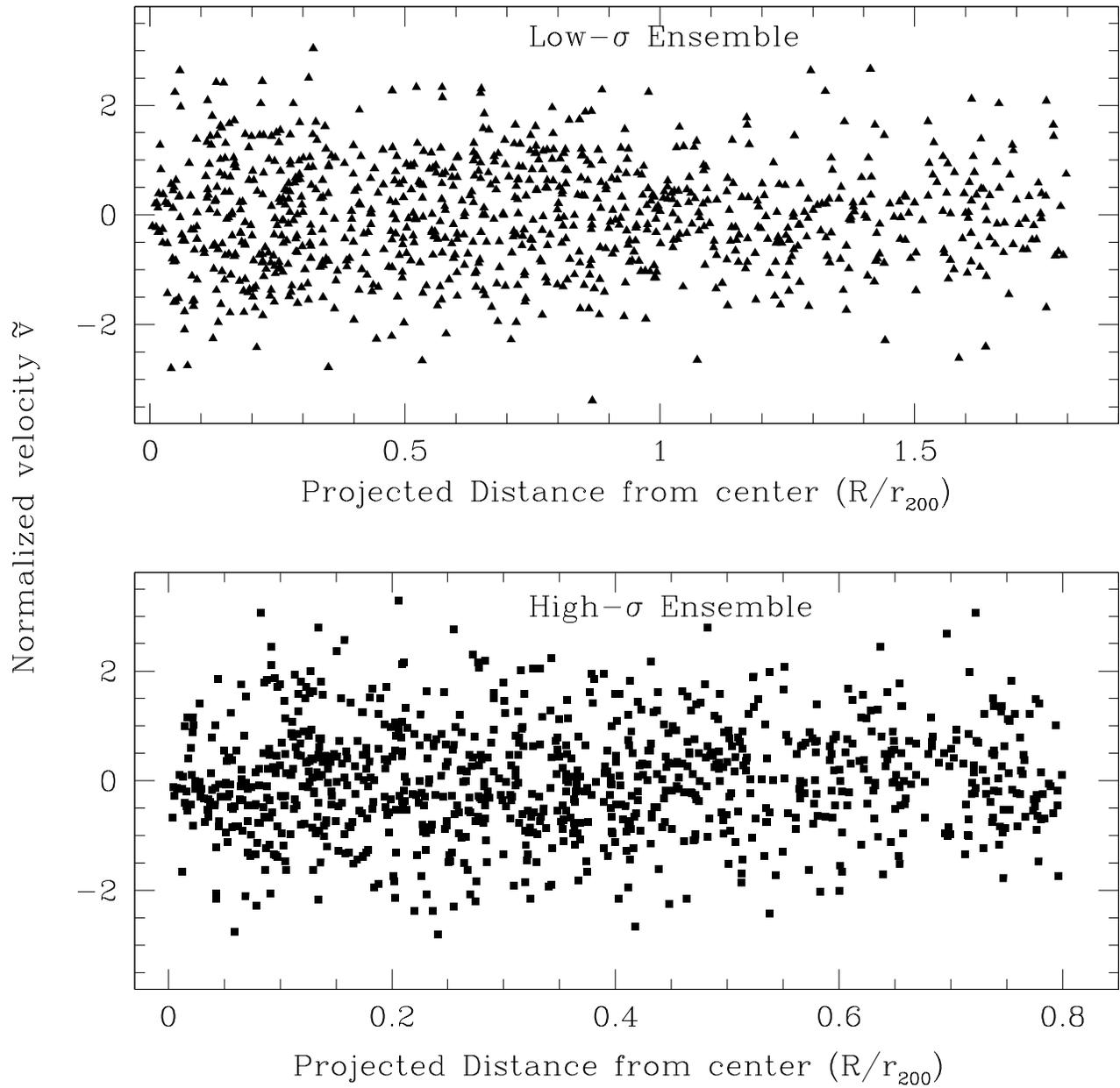}}
\figcaption{Phase space diagrams for the two ensembles. \label{fig:rvz}}
\end{center}
\end{figure*}

Our goal is to have an equal number of galaxies in the low- and
high-$\sigma$ ensembles. The closest we can come to this goal is to
split the systems at $\sigma = 470$ km s\m. Maximizing $N(\rlim)$
yields a low-$\sigma$ ensemble with 893 members complete to $\rlim =
1.8 \rtwo$, and a high-$\sigma$ ensemble with 945 members complete to
$\rlim = 0.8 \rtwo$. We therefore utilize 1838 (80\%) of the 2279
available member galaxy velocities in our data sample. The two
ensembles are shown in Figure \ref{fig:rvz}.

\subsection{Statistical Properties of the Ensembles}

\begin{figure*}
\begin{center}
\resizebox{7in}{!}{\includegraphics{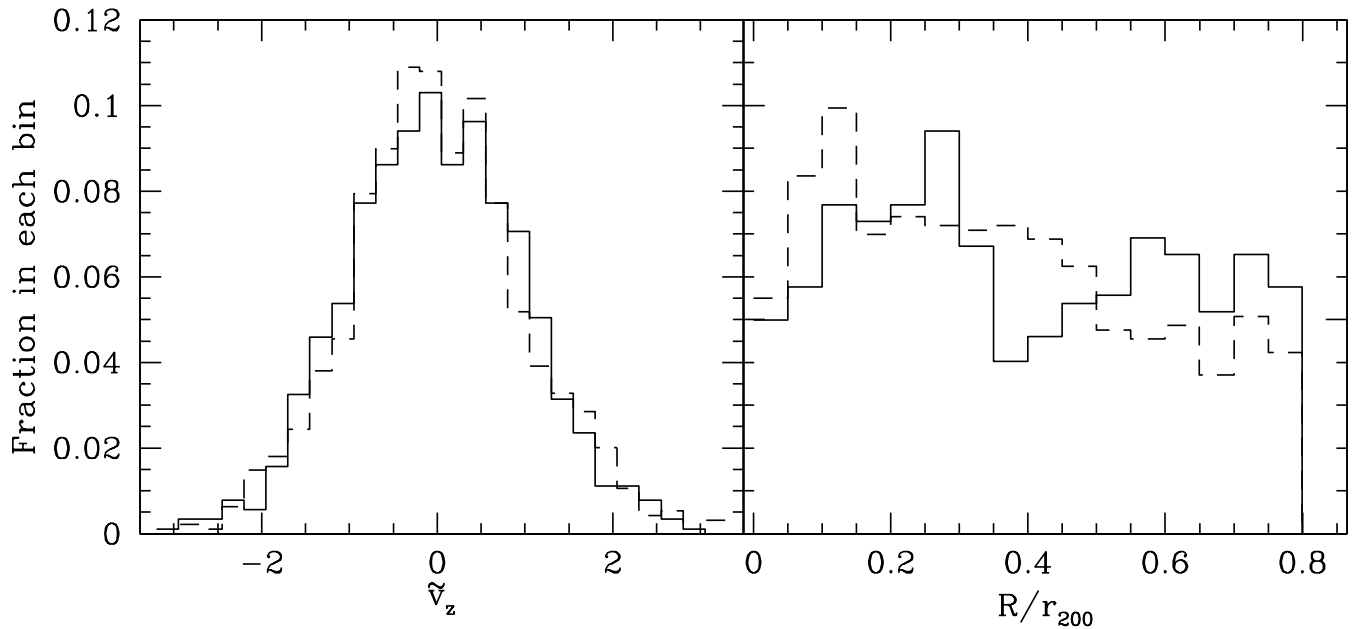}} \figcaption{
Comparison of the low-$\sigma$ sample (solid line) with the the
high-$\sigma$ sample (dashed line). Shown are the grand total velocity
histogram (\emph{left}) and the distribution of distances from the
cluster center (\emph{right}). For the latter plot, the low-$\sigma$
systems have been clipped at 0.8$\rtwo$.
 \label{fig:compare}}
\end{center}
\end{figure*}

Table \ref{tbl:props} shows the statistical properties of the final
dynamical sample. We use the two sample KS test \citep{NR} to evaluate
the consistency of the high- and low-$\sigma$ ensembles. The KS tests
show that the distributions of scaled velocities $\tilde{v}_z$ in the
two ensembles are consistent with each other. This is illustrated in
the velocity histograms in Figure \ref{fig:compare}.

On the other hand, the distribution of scaled distances $\tilde{R_i}$
differs in the two samples. According to the KS test, the probability
that the distributions are the same is less than $10^{-3}$.  Note that
for this comparison only, we truncated the low-$\sigma$ sample at
$\tilde{R} = 0.8$, because that is the limit to which the
high-$\sigma$ sample is complete. Figure \ref{fig:compare} also shows
the $\tilde{R}$ histogram. The distribution of the distances of
high-$\sigma$ galaxies varies more rapidly with $\tilde{R}$ than that
of the low-$\sigma$ sample.

Thus, although the galaxies in the high- and low-$\sigma$ ensembles
appear to have similar velocity distributions, their spatial
distributions differ slightly. In the next section we investigate
whether whether this difference translates into a difference in the
total matter distribution.

\section{Dynamics}
\label{sec:dynamics}

Here we discuss our technique for constraining the gravitational
potential and orbital distribution of a system of galaxies. We use
elements of \cite{Cuddeford}, \cite{Merritt93} and \cite{van00} in our
analysis. We add our own procedures for the faster computation of the
phase space integrals described below.

\subsection{Measurement Technique}

Our goal is to use measurements of galaxy positions and radial
velocities ($\mb{R},\mb{v_z}$) to constrain the shape of the matter
density. Our approach is Bayesian. For practical purposes,
``Bayesian'' means that our final result is a probability density
rather a set of numbers and associated errors. For example, in
addition to saying ``the mass is $M \pm \Delta M$ and the slope is $n
\pm \Delta n$,'' we draw curves of equal probability in $M - n$
space. These ``confidence regions'' will not necessarily be
elliptical, as would be the case for a straightforward linearized
$\chi^2$ fit.

Bayes's theorem specifies the method of calculating these confidence
regions. Let the vector $\mb{a}$ represent all the parameters we wish
to derive from the data. We seek the probability of $\mb{a}$ given the
data, which, according to Bayes's theorem, is\footnote{For simplicity,
we write the equations in this section in terms of $R$ and $v_z$,
rather than $\tilde{R}$ and $\tilde{v}_z$. }
\begin{equation}
p(\mb{a} | \mb{R},\mb{v_z}) = \frac{p(
\mb{a} )}{ p( \mb{R},\mb{v_z} ) } \prod_{i=1}^N p( R_i,v_{z,i} | \mb{a})
\label{eq:bayes}
\end{equation}
The function on the left is the joint probability distribution of the
unknown parameters $\mb{a}$, such that $\int p(\mb{a} | R,v_z ) d
\mb{a} = 1$.
On the right hand side:
\begin{enumerate}
\item $p( \mb{R},\mb{v_z} )$ is independent of $\mb{a}$. Because the
left hand side has to be normalized to unity, this probability is a
precisely calculable number.

\item $p( \mb{a})$ is the so-called Bayesian prior; it incorporates
our prior expectations regarding the value of $\mb{a}$. The priors are
detailed in \S\ref{sec:priors} below.

\item $p(R_i,v_{z,i} |\mb{a}) d R d v_{z}$ is the probability, given a
physical model $\mb{a}$, of observing a galaxy with a projected
distance from the center between $R_i$ and $R_i+ d R$, and with a
line-of-sight velocity between $v_{z,i}$ and $v_{z,i}+d v_{z}$.
\end{enumerate}
Through Bayes's theorem the desired confidence regions can be derived
by calculating the two-dimensional function $p(R,v_z|\mb{a})$ for all
$\mb{a}$ of interest, multiplying the result by the prior, and
normalizing the answer to unity.

To do this calculation we use the galaxy phase-space distribution
function (DF hereafter). The DF, a general way to describe the
dynamical state of the galaxies in a cluster, is sometimes written
$f(\mb{r},\mb{v})$, where $\mb{r}$ and $\mb{v}$ are the position and
velocity vectors, such that $f d^3 \mb{r}\, d^3 \mb{v}$ is the number
of galaxies within a small phase space volume element $d^3 \mb{r}\,
d^3 \mb{v}$ at $(\mb{r},\mb{v})$. If we knew the DF as a function of
our parameter set $\mb{a}$, it would be simple to write down the
probability of observing a galaxy with line-of-sight velocity $v_z$ at
a projected distance $R$ from the cluster center. We begin by
referring to this probability as $p^\prime(R,v_z|\mb{a})$.

Suppose we conduct a complete redshift survey of an isolated spherical
cluster complete to a limiting projected distance $R_\mr{lim}$ from
the center. Then the probability of observing a given galaxy with
radius $R$ and velocity $v_{z}$ is
\begin{eqnarray}
p^\prime(R,v_z|& &\mb{a}) = \frac{2 \pi R}{N_0} \times  \nonumber \\
& & \int \int \int f\left[ \mb{a}(R,z,v_R,v_\phi,v_z) 
\right] dv_R\, dv_\phi \, d z
\label{eq:probdef}
\end{eqnarray}
where the $R$ and $\phi$ denote polar coordinates in the plane of
the sky, and $z$ is the line of sight.  The normalization $N_0$ is
the total number of galaxies within the limits of the survey, such
that
\begin{equation}
\int_{0}^{R_\mr{lim}} \int_{-\infty}^{\infty} 
p^\prime(R,v_z|\mb{a}) dR \, dv_z = 1.
\label{eq:pnorm}
\end{equation}
The chief difficulty lies in calculating the DF as a function of the
dark matter potential and the galaxy orbital distribution. In other
words, we must find the functions $\mb{a}(R,z,v_R,v_\phi,v_z)$.  The
probability $p'$ represents an idealized observation of a cluster in
the absence of contaminants (``interlopers''); the true probability
$p$, which takes interlopers into account, is calculated below
(\S\ref{sec:interlopers}).

We can begin by embedding this DF into our spherical gravitational
potential $\Phi(r)$. Because $\Phi$ is a negative quantity by
convention, it is helpful to redefine $\Psi \equiv
\Phi(\infty)-\Phi(r)$. Neglecting encounters, we assume that the only
force on a galaxy is the Newtownian gravitational interaction
$\dot{\mb{v}} = \grad \Psi$. If we could write the DF in terms of the
acceleration $\dot{\mb{v}}$ and the potential $\Psi$, we would have
the required link between the phase space variables and the
gravitational potential. Such an equation would connect the
observations to the matter distribution.

The Jeans theorem
\citep{BinneyTremaine87}
provides this crucial link.  The theorem guarantees that in a
steady-state spherical system the DF is a function of only two
integrals of motion: the negative energy $\ene \equiv \Psi - v^2/2$,
where $v^2 = \mb{v} \cdot \mb{v}$, and the square of the angular
momentum $L^2$. Thus we can write
\begin{equation}
f = f( \ene, L^2 ) = f \left[ \Psi(r) - v^2/2 \, , |\mb{r} \cross \mb{v}|^2 
\right]
\label{eq:df}
\end{equation}
Newton's law and the Jeans theorem provide all the information we need
to calculate the probability of observing a data point $(R,v_z)$ given
a DF and potential ($f$,$\Psi$) through equations
(\ref{eq:probdef})-(\ref{eq:df}). The Appendix contains the full
expression for $p^\prime(R,v_z|\mb{a})$ in spherical coordinates.

Once we calculate the probability of the data given a specific model,
we can use Bayes's theorem to calculate constraints on the set of
model parameters.

\subsection{Mass Model}

To begin we consider the parameters of the mass profile. Many forms
exist in the literature, but the ideal profile for our purposes should
fulfill the following criteria, in decreasing order of importance: (1)
It should be general enough to have the inner matter density slope as
a free parameter; (2) it should include well-studied profiles as
special cases; and (3) it should generate a density and gravitational
potential in terms of elementary functions. This last property would
make phase space distribution analysis faster and more
straightforward.

We use a mass profile that fulfills all the three criteria: 
\begin{equation}
M(r) = M_0  \left( \frac{r}{r+r_0} \right)^{3-n},
\label{eq:massmodel}
\end{equation}
where $M_0$ is the total mass of the system, $r_0$ is the
characteristic radius, and $n$ is the inner slope of the matter
density. This model is sometimes referred to as the ``gamma-model''
(not to be confused with $\gamma$ below) in stellar-dynamical
literature \citep{Dehnen93,Tremaine94}. The density and potential pair
generated by this mass profile are
\begin{eqnarray}
\label{eq:mainpot}
\rho(r) & = & \frac{M_0 r_0 (3-n)}{4 \pi} r^{-n} (r_0+r)^{n-4} \\
\Phi(r) & = & -\frac{G M_0}{r_0 (2-n)} \left[1 - \left( \frac{r}{r+r_0} \right)^{2-n} \right]
\end{eqnarray}
The above profile includes the well-known \cite{Hernquist90} and
\cite{Jaffe83} models, which are generated by the $n=1$ and $n=2$
cases, respectively. The Hernquist profile is known to match the light
and matter distribution in elliptical galaxies as well as in groups
and clusters \citep{Mahdavi99,Rines01}. The formulation above is a
generalization with arbitrary inner slope $n$. Note that the NFW model
$\rho_\mathrm{NFW} \propto r^{-1} (1+r)^{-2}$ has infinite mass and is
not a subset of the models we consider. Using a method completely
different from ours, \cite{Rines03} find that samples as large $10^4$
galaxies cannot distinguish between the NFW and Hernquist mass
distributions in the infall regions of clusters ($R > 2$ Mpc).

As a further boon, the generalized potential is invertible
analytically:
\begin{equation}
r(\Phi) = \frac{r_0}{1 - [1 + (n-2) \Phi r_0 / G M_0 ]^{1/(2-n)}} ,
\end{equation}
further simplifying the calculation of phase space densities described
below.

In summary, we begin with a three dimensional parameter set $\mb{a}$,
containing the inner slope, the transition radius, and the
normalization of the matter density. Three more parameters not
directly related to the matter density are also required. They are the
slope of the galaxy density at infinity, $\gamma$, the velocity
anisotropy, $\beta$, and the interloper fraction, $P_I$.

\subsection{DF model}
\label{sec:df}

Next we examine the form for the DF, $f(\ene,L^2)$.  It is virtually a
guarantee that given a potential $\Psi$, many quite different $f$ will
describe the same data equally well. We therefore use a very simple
DF---a scale free, separable function of the energy and angular
momentum \citep{Fricke52}:
\begin{equation}
f(\ene,L^2) = f_0 \ene^{\alpha-1/2} L^{-2 \beta}.
\end{equation}
As we show below, this well-studied DF adequately describes the
cluster and group data.

This DF yields a unique galaxy density profile $\nu$ (distinct from
the matter density profile $\rho$). To calculate a galaxy density
profile, we note that the integral of this particular DF over all
velocities gives the particle space density; see the Appendix for a
detailed calculation of the integral:
\begin{eqnarray}
\nu(r) & = & \int_{v^2 \le 2 \psi} f(\ene,L^2) d^3 \mb{v} 
\label {eq:dfint} \\
& \propto & r^{-2 \beta} \Psi(r)^{\alpha-\beta+1}.
\label{eq:nu}
\end{eqnarray}

This $\beta$ parameter is equal to the orbital anisotropy of the
galaxies:
\begin{equation}
\beta = 1 - \frac{\sigma_t^2}{\sigma_r^2}
\end{equation}
In other words, $\sqrt{1-\beta}$ is the ratio of the velocity
dispersion in the tangential direction to the velocity dispersion in
the radial direction. From both the equations above it is evident that
$\beta < 1$. If that were not the case, (1) the galaxy mass $\int
r^2 \nu(r) d \,r$ would diverge at $r=0$, and (2) one of the velocity
dispersions $\sigma_t$ and $\sigma_r$ would have to be imaginary.

The $\alpha$ and $\beta$ parameters also describe the slope of the
galaxy density at infinity. For example, if $\Psi(r) \propto 1/r$ at
large radii, then $\nu(r) \propto r^{-\gamma}$, with
\begin{equation}
\label{eq:gamma}
\gamma = \alpha+\beta+1.
\end{equation}
Henceforth, without loss of generality, we will discuss the energy
slope of the DF solely in terms of $\gamma$. As $\gamma$ increases, the cluster contains fewer galaxies with large
kinetic energies. This behavior occurs because galaxies with energy
$\ene \sim \Phi$ begin to outnumber galaxies with $\ene \approx 0$
(see equation \ref{eq:df}). As a result the galaxy distribution is
less extended than it would be with small $\gamma$. The minimum
acceptable value is $\gamma = 4$, because the mass models we consider
(equation \ref{eq:mainpot}) decline asymptotically as $r^{-4}$.

The inner slope of the galaxy density $n_g$ can be different from the
slope of the matter density $n$:
\begin{equation}
n_g  = \left\{
\begin{array}{lr}
2 \beta  & (n < 2) \\
2 \beta + (n-2)(\gamma-2\beta)& (n > 2) \\
\end{array}
\right.
\end{equation}
The reason for the split at $n=2$ is that the potential $\Phi$ becomes
singular for $n \ge 2$, and through equation (\ref{eq:nu}), gives a
contribution to the inner slope of the galaxy density. For $n < 2$,
the potential is constant at $r = 0$, and there is no such
contribution. A corollary of this property is the requirement
\begin{equation}
n > n_g.
\end{equation}
This statement specifies that the galaxy density may never exceed the
total matter density. It implies, importantly, that for constant
density cores $n=0$, no radially anisotropic models ($\beta > 0$) are
viable. This requirement is not unique to our model; a large number of
other radially anisotropic models, with substantially different DFs
and anisotropy profiles $\beta(r)$, are also incompatible with
constant-density cores \citep{van00,Cuddeford}.

Up to this point we have a five-dimensional parameter space $\mb{a} =
(M_0,n,r_0,\gamma,\beta)$. One final dimension is necessary, the
interloper fraction.

\subsection{Interlopers}
\label{sec:interlopers}

So far we have assumed that the clusters and groups we observe exist
in isolation, free from any contamination from background and
foreground galaxies. In reality, however, even after velocity clipping
(see \S\ref{sec:membership}), a small percentage of the galaxies are
unrelated objects projected into the phase space volume of
interest. We wish to estimate this percentage---the interloper
fraction $P_I$---from the data.

We follow the elegant method of \cite{van00} in treating the
interlopers statistically. This method assumes that interlopers land
randomly in the group or cluster phase space volume. Thus every galaxy
we observe has a finite probability $P_I$ of being an unrelated
interloper. The probability that a galaxy with velocity $v_z$ and
projected distance $R$ is included in our survey is then
\begin{equation}
p(R_i,v_{z,i}|\mb{a}) = (1-P_I) p^\prime(R_i,v_{z,i}|\mb{a}) + 
\frac {2 P_I R_i}{R_\mr{lim}^2 | v_\mr{lim} |} 
\label{eq:pfinal}
\end{equation}
Where $|v_\mr{lim}|$ is the maximum observed line of sight
velocity. It is easy to verify that the above expression meets a
normalization criterion similar to equation (\ref{eq:pnorm}) as long
$p^\prime(R,v_z|\mb{a}) \rightarrow 0$ as $v_z \rightarrow
|v_\mr{lim}|$:
\begin{equation}
\int_{0}^{R_\mr{lim}} \int_{-|v_\mr{lim}|}^{|v_\mr{lim}|}
p(R,v_z|\mb{a}) dR \, dv_z = 1.
\end{equation}

We can now write down the final expression for the six-dimensional
function $p(\mb{a} | \mb{R},{v_z})$. The full expression is too long
to be listed here and is derived in detail within the Appendix. The
derivation consists of a straightforward if laborious combination of
equations (\ref{eq:probdef})~-~(\ref{eq:pfinal}). The main
complications occur because we observe a spherical object in
cylindrical coordinates. 

We henceforth refer to the joint probability distribution of
observables for a given measurement simply as
$p(\tilde{R},\tilde{v}_z)$, where without loss of generality we have
made the substitution $R \rightarrow R/\rtwo \equiv \tilde{R}$ and
$\tilde{v}_z \rightarrow v_z/\sigma_z \equiv \tilde{v}_z$. 

\subsection{Reparameterization and Bayesian Priors}
\label{sec:priors}

The six parameters of our model are
($n$,$M_0$,$\gamma$,$r_0$,$\beta$,$P_I$).  Some of these parameters
can be redefined without loss of generality to yield more compact
joint confidence regions. For example, we find that jointly
constraining $n$, $M_0$ and $r_0$ as defined in equation
(\ref{eq:massmodel}) yields a high degree of correlation among all
three parameters.  However, by switching to the system 
\begin{eqnarray}
r_{5/2} & \equiv & \frac{3 r_0}{5 - 2 n}, \\ M_{200} & \equiv &
M(\rtwo),
\end{eqnarray}
we significantly reduce the overall extent of the confidence
regions. The new radius scaling $r_{5/2}$ is defined such that
\begin{equation}
\left. \dlndln{\rho}{r} \right|_{r=r_{5/2}} = -\frac{5}{2}
\end{equation}
In other words, $r_{5/2}$ is the location at which the slope of the
matter density is 5/2. Similarly, $M_{200}$ is the total mass
contained within $r_{200}$. 

The fact that we analyze ensembles of clusters has implications for
the measurement of $r_{5/2}$ and $M_{200}$. To create the ensembles,
it was necessary to scale the radii and line-of-sight velocities by
$\rtwo$ and $\sigma$, respectively (see \S\ref{sec:reject}).  Because of
this scaling, the mass and characteristic radius of the ensembles are
normalized as well. As a result, our method constrains the
dimensionless quantity 
\begin{equation}
\tilde{M}_{200} \equiv \frac{G M_{200}}{r_0 \sigma^2}
\end{equation}
 for each of the two the ensembles.  Similarly, the quantity
\begin{equation}
\tilde{r}_{5/2} \equiv r_{5/2} / r_{200},
\end{equation}
rather than $r_{5/2}$ alone, is constrained by our method.

A redefinition of the anisotropy is also appropriate. Because the
allowable range in $\beta$ is [$-\infty$,1], we define, similarly to
\cite{Wilkinson02},
\begin{equation}
\tilde{\beta} \equiv -\ln{(1 - \beta)}.
\end{equation} 
This redefinition maps the allowable range from [$-\infty$,1] into
[$-\infty,\infty$], and ensures that the parameter space has equal
prior probability density in the radially and tangentially anisotropic
regimes.

The final parameter set is
($n$,$\tilde{M}_{200}$,$\tilde{r}_{5/2}$,$\gamma$,
$\tilde{\beta}$,$P_I$). In our analysis, we take the Bayesian priors
$p(\mb{a})$ (equation \ref{eq:bayes}) for these parameters to be
uniform.

\subsection{Computation Technique}
\label{sec:master}

Here we describe the numerical procedure used to calculate our
probabilities.  Even though $p(\tilde{R},\tilde{v}_z)$ has six
dimensions, we limit ourselves to constraining five at any given
time. As we explain in \S\ref{sec:freen}, for free $n$ our data can
place only a lower, and not an upper limit on the energy slope
$\gamma$, and therefore $p(\tilde{R},\tilde{v}_z)$ cannot be
normalized properly as a function of $\gamma$. On the other hand, if
we fix $n$, then we can easily constrain $\gamma$. For this reason, we
conduct the entire procedure described below several times: (a) once
with $n$ fixed and $\gamma$ free (\S\ref{sec:freeg}), and (b) many
times with $n$ free and $\gamma$ ranging over several discrete values
(\S\ref{sec:freen}).

We use a computation technique that yields an estimate of
$p(\tilde{R},\tilde{v}_{z})$ on an grid with $m^5$ elements:

\begin{enumerate}

\item \emph{Tabulation}. There are two scaling parameters, the
transition radius $r_{5/2}$ and the mass normalization $M_{200}$. We
define two dimensionless parameters, $R^\prime \equiv
\tilde{R}/\tilde{r}_{5/2}$, and $v^{\prime 2} \equiv \tilde{v}_z^2 r_0
/ (G M_{200})$.  Then we tabulate $p(R^\prime, v^\prime | \mb{a})$ on
a refined $M \times M$ grid for different values of $\tilde{\beta}$,
and $n$. This step therefore requires $m^2 M^2$ evaluations of the
triple integral given in equation (\ref{eq:probdef}). This phase
occupies roughly 25\% of total computation time.

\item \emph{Interpolation}. For each of the $N_0$ data points
$(\tilde{R}_i,\tilde{v}_{z,i})$, we use cubic spline interpolation to
calculate the probability density as a function of $\tilde{\beta}$,
$\tilde{M}_{200}$, $\tilde{r}_{5/2}$, and either $n$ or $\gamma$. This
step requires $N_0 m^4$ interpolations, and takes up roughly 70\% of the
total computation time. Because of the linearity of the interloper
parameter in equation (\ref{eq:pfinal}), calculating the probability
density as a function of $P_I$ takes up a trivial amount of time in
comparison to the $N_0 m^4$ interpolations.

\item \emph{Projection}. To display the 5-dimensional probability
density, we produce two-dimensional realizations of it. This means
creating $5 \times 4 / 2 = 10$ separate probability distributions,
each representing a summing (or ``marginalization'') of $p(\mb{a} |
\mb{R},\mb{v_z})$ along a cube orthogonal to the other 9. For example,
to obtain the joint probability distribution of the mass normalization
$\tilde{M}_{200}$ and inner slope $n$, we use
\begin{equation}
\int \int \int p(\tilde{M}_{200},n,\tilde{r}_{5/2},\tilde{\beta},P_I)
d \tilde{r}_{5/2} \, d \tilde{\beta} \, d P_I.
\end{equation}
This step is conducted simultaneously with step (2) and takes up
a negligible amount of the total computation time.

\item \emph{Goodness of fit.} To check the goodness-of-fit of the
physical model, we bin the data in two dimensions and use $\chi^2$
tests to compare the number of galaxies in each bin to the number
predicted by the most likely model. We bin the data only to check the
goodness of fit, not to derive constraints on the parameters. We
calculate the quantity
\begin{equation}
X^2 \equiv  \sum_{j=1}^k \frac{(N_j - p_j N)^2}{N_j},
\end{equation}
where $k$ is the total number of bins, $N_j$ is the number of galaxies
observed in bin $j$, and $p_j$ is the integrated probability of the
best fit model in that bin. The quantity in the denominator is the
Poisson variance in each bin. 

It can be shown that for our case $X^2$ is bounded above by a
$\chi^2_{k-1}$ distribution and below by a $\chi^2_{k-s-1}$
distribution \citep{Lupton}, where $s=5$ is the number of parameters
we fit. Therefore, if $1-q$ is confidence with which we can reject the
hypothesis that the data is consistent without our model, then $q$
must lie between $\Gamma[(k-s-1)/2,X^2/2]$ and $\Gamma[(k-1)/2,
X^2/2]$, where $\Gamma[a,x]$ is the incomplete gamma function.  We
always use the value that yields the smaller $q$.

We create rectangular bins in $R,v_z$ space so that they each contain
30 galaxies and so that $\sum p_j = 1$. Let $t$ be the largest integer
smaller than or equal to $\sqrt{N_0/30}$. There are $t$ strips in the
radial direction, each with 30$t$ galaxies. The first $t-1$ radial
strips are divided into $t$ bins each containing 30 galaxies; the last
radial strip is divided into $t-1$ bins each containing 30 galaxies,
and a final bin containing the remaining galaxies. Therefore, the
total number of bins is $k = t^2$.

\end{enumerate}

For our final calculations, we use $M=50$ and $m=40$.

\subsection{Simulations}
\label{sec:sims}

The ideal procedure to test the validity of our technique would be to
create thousands of Monte Carlo realizations of various clusters with
different values of $\beta$, $\gamma$, $r_0$, $M_{200}$, and $P_I$,
and to conduct our analysis separately for each realization. Then we
could compute the fraction of simulations in which the most likely
parameters closely match the input parameters. Unfortunately, such a
test would be prohibitively expensive, requiring many months of CPU
time.

As a compromise, we conduct four tests in which we attempt to model
simulated ensembles. In these simulations, we draw 893 members from
the $p(\tilde{R},\tilde{v}_z)$ for a set of parameters given \emph{a
priori}. The procedure for drawing deviates distributed according to
$p(\tilde{R},\tilde{v}_z)$ is the rejection method described in
\cite{NR}. The rejection method uses a comparison function that is
everywhere greater than $p(\tilde{R},\tilde{v}_z)$ and has an analytic
cumulative distribution. A large number of deviates are drawn from the
comparison function, and those that lie beneath
$p(\tilde{R},\tilde{v}_z)$ have the desired probability
distribution. We extend the \cite{NR} technique to two dimensions by
using a constant comparison function in $R,\tilde{v}_z$.  Finally, we
maximize the Bayesian likelihood by conducting the entire procedure
described in the previous section. The simulated data are drawn from
models with $\gamma = 8$. We first analyze the simulations assuming
$\gamma=8$, and then we consider the effects of varying $\gamma$.

\begin{figure*}
\resizebox{7in}{!}{\includegraphics{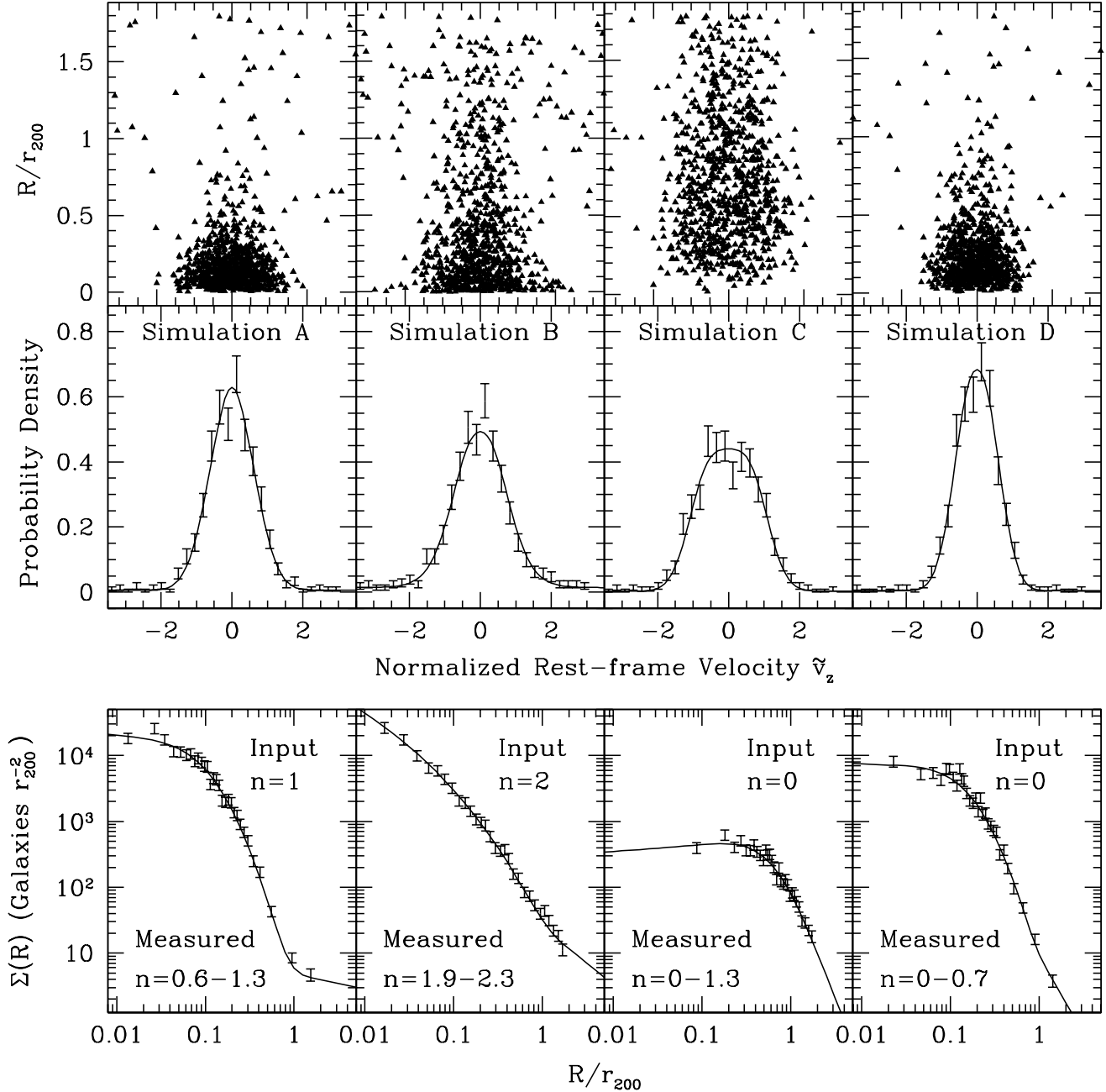}} \figcaption{Simulations
of the technique: (\emph{top}) phase space distributions generated
using the four parameter sets in Table \protect\ref{tbl:sims};
(\emph{middle}) the grand total velocity histogram $N(\tilde{v}_z)$;
(\emph{bottom}) the galaxy surface density. The solid lines show
predictions (not fits) of the maximum-likelihood analysis of unbinned
distance and velocity data. 
The ``breaks'' in $\Sigma(R)$
around $R \sim r_{200}$ in simulations A and B are due to contamination
by the interloper population $P_I$. \label{fig:sims} }
\end{figure*}

Table \ref{tbl:sims} lists the parameters used for the simulations,
and Figure \ref{fig:sims} shows their properties. The phase space
distribution generated from each parameter set appears together
with the grand total velocity histogram,
\begin{equation}
N(\tilde{v}_z) \equiv \int_{0}^{\tilde{R}_\mr{lim}}
p(\tilde{R},\tilde{v}_z) d\tilde{R}
\label{eq:nvz}
\end{equation}
and the surface number density
\begin{equation}
\Sigma(\tilde{R}) \equiv N_0 \int_{-\infty}^{\infty}
p(\tilde{R},\tilde{v}_z) d \tilde{v}_z / ( 2 \pi \tilde{R} ),
\label{eq:sigma}
\end{equation}
where $N_0$ is the total number of galaxies in the sample. In
these figures we also show the most likely models for the above two
observables. Note that these curves are not fits, but predictions of
the unbinned maximum likelihood technique described in the previous
section.

According to the simulations, our estimation technique is able to tell
apart fundamentally different models with similar binned
statistics. For example, simulation A and D have similar
$N(\tilde{v}_z)$ and $\Sigma(\tilde{R})$, even though A was generated
with $n=1$ and D was generated with $n=0$. Because we analyze unbinned
velocities and radii, our technique yields a 95\% confidence region $n
= $0.6--1.3 for A and $n = $0--0.7 for D.  The unbinned estimator is
able to distinguish the parent distributions of the two data sets.

Another feature of the simulations is that they are more powerful when
$n$ is steep---that is, the slope of the matter density is large. In
this case our method yields a fairly accurate measurement of $n$. When
$n$ is close to 0---that is, the matter density has a flat core---then
the measurement method yields a wider range in $n$. Nevertheless, this
range for both models C and D includes the correct $n=0$ solution.

We can also use the simulations to test the validity of our numerical
integration. The integrals of $N(\tilde{v}_z)$ and $2 \pi \tilde{R}
\Sigma(\tilde{R})/N_0$ should equal 1. For the four
simulations, we find that $\int N(\tilde{v}_z) = 0.9976, 0.9880,
0.9998, 0.9987$, and $\int 2 \pi \tilde{R} \Sigma(\tilde{R}) / N_0 =
0.9987, 0.9972, 1.0001,0.9995$. Thus in most cases our accuracy is
better than 0.5\%.

\begin{figure*}
\resizebox{7in}{!}{\includegraphics{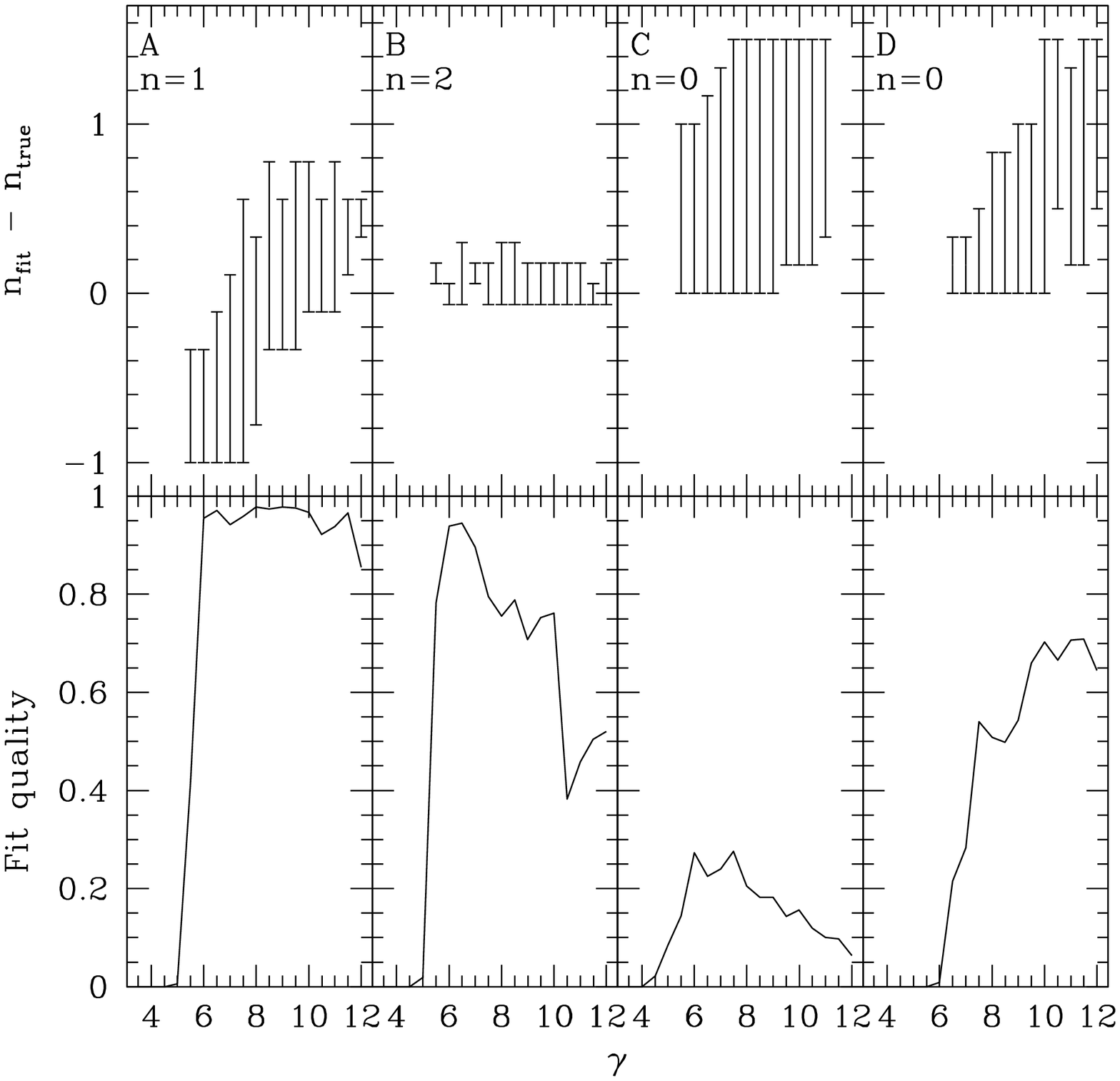}} \figcaption{ Testing
the robustness of the simulations as a function of $\gamma$:
(\emph{top}) the 95\% confidence regions on the best-fit $n$ minus the
true $n$ for each simulation; (\emph{bottom}) the quality of fit as a
function of $\gamma$.
\label{fig:gamma}}
\end{figure*}

Finally, it is useful to examine how sensitive the measurement
technique is to variations in $\gamma$. In analyzing our simulations,
we have set $\gamma=8$, i.e., we have fixed it at the value used to
generate the data. Now we reanalyze the data assuming different values
of $\gamma$, from 4 to 12.  We rerun our analysis procedure,
calculating the quality of fit $q$ and the 95\% marginalized
confidence region in $n$ as a function of $\gamma$. The results are
shown in Figure \ref{fig:gamma}.  We find that as long as $q > 0.1$,
i.e., the fit is statistically acceptable, then the derived confidence
regions $n_\mathrm{fit}$ are close to the true $n$ for most values of
$\gamma$.  In general, choosing the \emph{smallest value} of $\gamma$
that gives $q > 0.1$ results in the most accurate confidence intervals
on $n$. Choosing larger values of $\gamma$ yields confidence regions
$n_\mr{fit}$ that are slightly different from the correct values.

While not exhaustive, these simulations suggest that our measurement
technique is trustworthy. The 95\% confidence contours derived using
our method should include the correct value of each parameter if the
general model is admissible (i.e., $q \gtrsim 0.1$), with the caveat
that the smallest value of $\gamma$ that yields a good fit should be
used. These confidence regions ought to be compact for large values of
the density slope $n$, and broad-valued for flat halos with small $n$.

\section{Results}
\label{sec:discuss}

Here we describe our constraints on the inner matter density slope
$n$, the mass normalization $\tilde{M}_{200}$, the transition radius
$\tilde{r}_{5/2}$, the velocity anisotropy $\tilde{\beta}$, and the
interloper fraction $P_I$.  It bears repeating that our constraints
result from maximization of the likelihood function described above,
using the full unbinned data set. The binned profiles we show are for
illustrative purposes and do not indicate ordinary $\chi^2$ fitting.

\subsection{Models with Constant-Density Cores}
\label{sec:freeg}

First we investigate the possibility that the ensembles have
constant-density cores ($n=0$). In \S\ref{sec:df} we showed that such
cores do not support radially anisotropic models; we are limited to
$\beta \le 0$. Therefore the minimization occurs over five dimensions:
($\tilde{\beta}$,$\gamma$,$\tilde{r}_{5/2}$,$\tilde{M}_{200}$,$P_I$).

\begin{figure*}
\resizebox{7in}{!}{\includegraphics{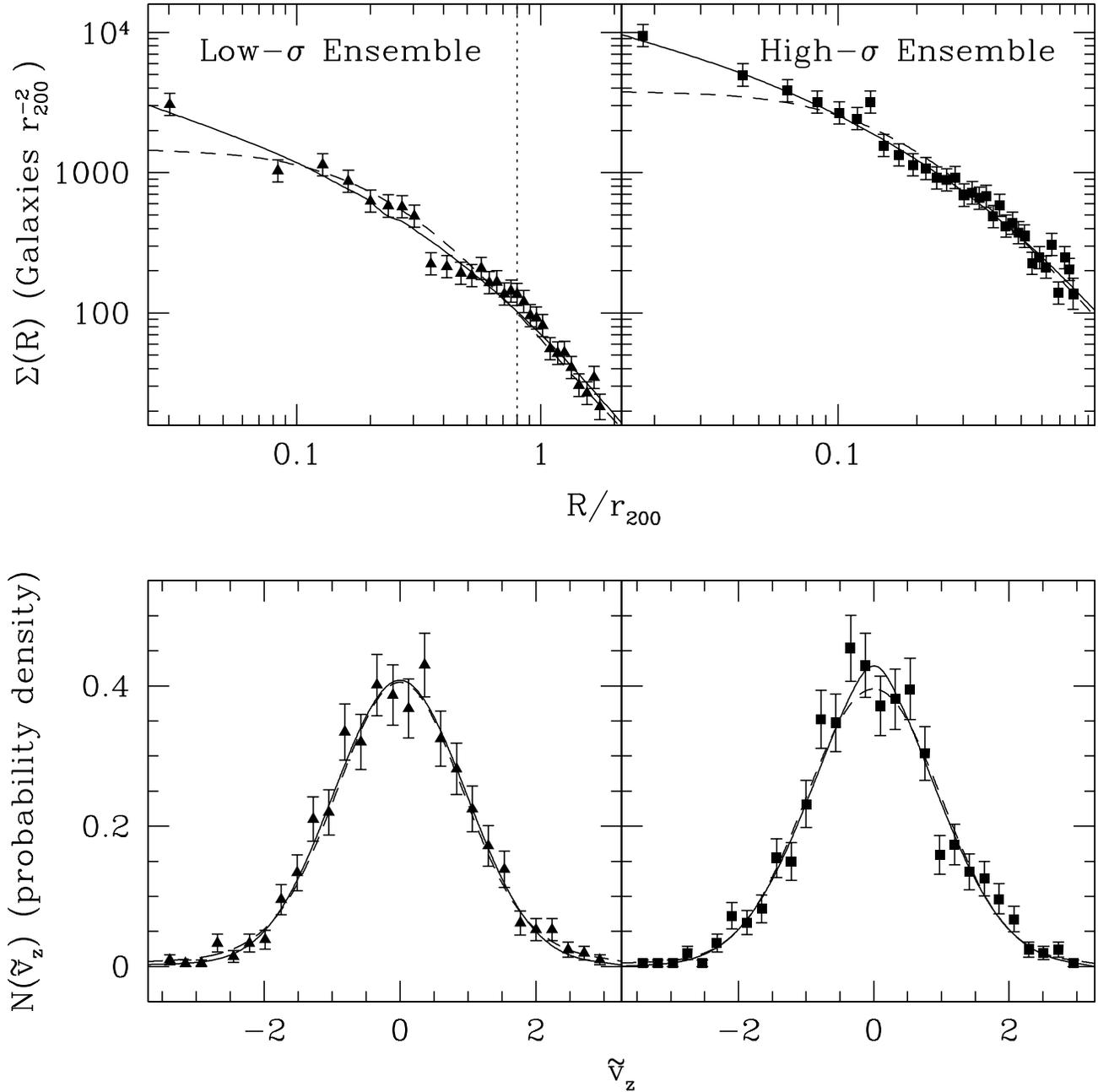}}
\figcaption{Predictions (not fits) of the maximum-likelihood analysis
of unbinned distance and velocity data. Shown are the surface number
density (\emph{top}) and the grand total velocity histogram
(\emph{bottom}).  The solid line represents the most likely model
overall, whereas the dashed line shows the most likely model when the
inner slope of the matter density is forced to equal 0. The latter
model is rejected with 99.7\% or better confidence. The dotted
vertical line represent the completeness limit of the high-$\sigma$
ensemble. \label{fig:results}}
\end{figure*}

The dotted line in figure \ref{fig:results} shows the results of the
minimization. We show the most likely grand total velocity histogram
and galaxy surface density as given by equations
(\ref{eq:nvz})-(\ref{eq:sigma}). 

We find that matter profiles with constant-density cores do not
produce a galaxy density profile as steep as our data.  Although the
grand-total velocity histogram $N(\tilde{v}_z)$ seems accurately
reproduced, the model's galaxy density in the innermost regions is too
small. The constant density models are disfavored with $q = 0.003$ for
the low-$\sigma$ ensemble, and $ 10^{-5}$ for the high-$\sigma$
ensemble. 

Note that the inconsistency with the galaxy surface density $\Sigma$
is not the only discrepancy between the data and the $n=0$ model. Our
$\chi^2$ goodness-of-fit method described above shows that the $n=0$
models predict too many high-velocity galaxies at intermediate
radii. This discrepancy is not easily visualized through the plots of
$N(\tilde{v}_z)$ or $\Sigma(\tilde{R})$.

Just to be sure that the $n = 0$ models are a poor fit, we also try
relaxing the requirement that the orbits not be radially anisotropic.
We fit the full range of negative to positive $\beta$, and find that
$\beta \approx 0.25, \gamma \approx 4$ maximize the likelihood for
both ensembles. These values are formally unphysical, yielding a
galaxy density that is steeper than the $n = 0$ total matter density
near the central regions. However, even these fits are rejected with
$q = 0.02$ for groups and $q = 0.003$ for clusters.

It is interesting to note (Figure \ref{fig:results}) that the
inconsistency occurs near $r = 0$, and not at larger radii. If our
choice of mass model (equation \ref{eq:mainpot}) were the cause for
the poor fit, we would expect the inconsistency to occur at the larger
radii, because that is where we impose a $\rho \propto r^{-4}$
behavior. We conclude that models with a constant density core ($n =
0$) are at best barely consistent with the data.

\subsection{Models with Fixed $\gamma$}
\label{sec:freen}

Now we allow the inner slope of the mass density $n$ to vary,
maximizing $p(\tilde{R},\tilde{v}_z|\mb{a})$ for several different
values of $\gamma$, the slope of the galaxy density at large
radii. The jointly constrained parameter set is therefore
($n$,$\tilde{\beta}$,$\tilde{r}_{5/2}$, $\tilde{M}_{200}$,$P_I$).  As
described in \S \ref{sec:df}, $\gamma$ is also related to the slope of
the energy term in the distribution function (equation
\ref{eq:df}). The larger the value of $\gamma$, the more peaked the
distribution of galaxies with small kinetic energies.

Our data can provide a lower, but not an upper limit on $\gamma$. For
both the low- and the high-$\sigma$ ensembles, the quality of the fit
increases with $\gamma$, asymptotically approaching a fixed value as
$\gamma \rightarrow \infty$. Our simulations (\S\ref{sec:sims}) show a
similar effect, and suggest that the smallest value of $\gamma$ that
gives an acceptable fit is likely to give the most accurate results.
Therefore, instead of constraining $\gamma$ continuously, we list the
minimum value of $\gamma$ that yields a quality of fit $q > 0.1$. This
constraint corresponds to $\gamma=8$ for the low-$\sigma$ and
$\gamma=12$ for the high-$\sigma$ ensemble.

\begin{figure*}
\begin{center}
\resizebox{7in}{!}{\includegraphics{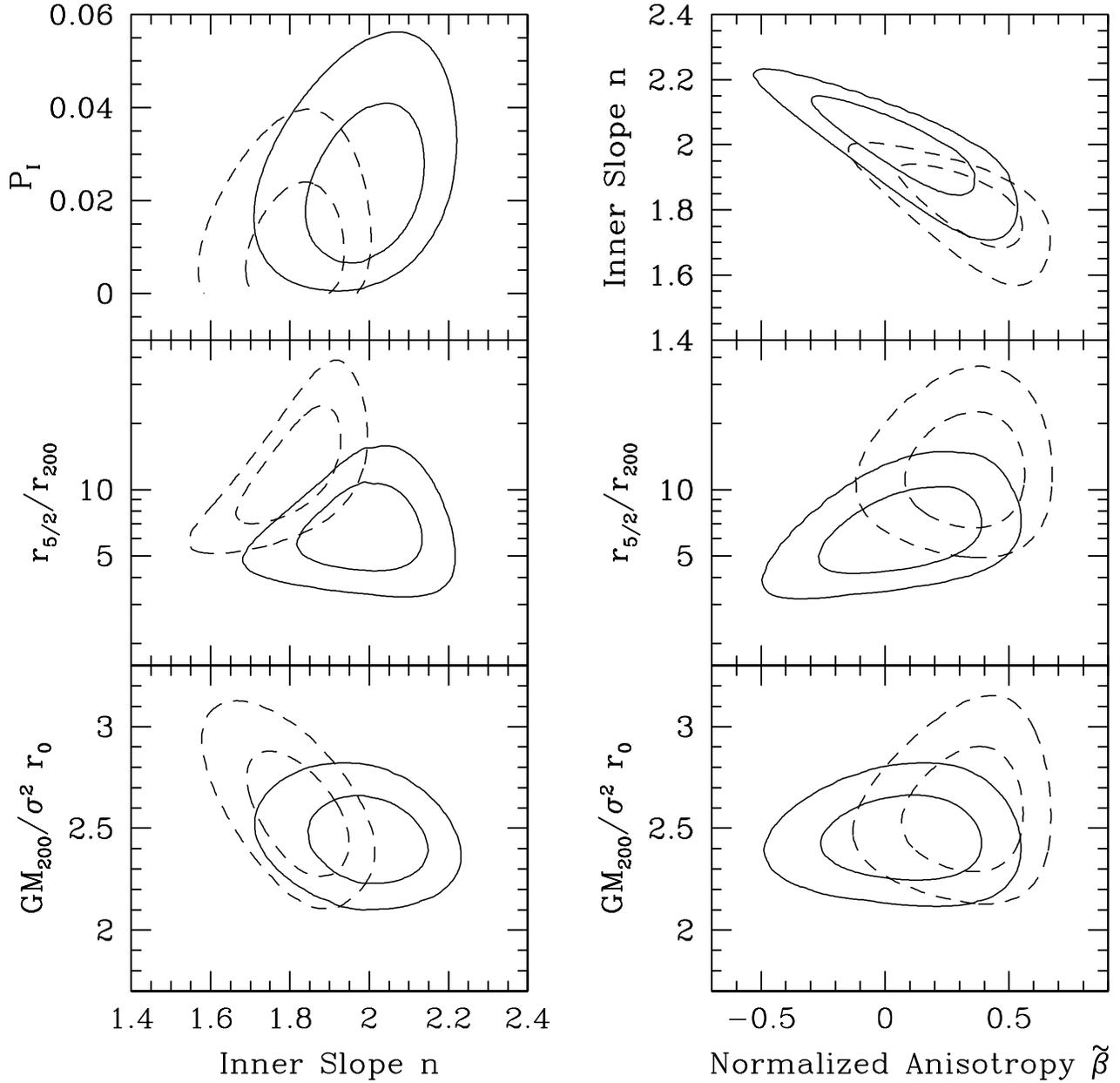}} \figcaption{Results
of the likelihood maximization for free $n$. Solid contours represent
the low-$\sigma$ ensemble and dashed contours represent the
high-$\sigma$ ensemble. Shown are marginalized joint 68\% and 95\%
probability contours in the matter density slope $n$, normalized
anisotropy $\tilde{\beta}$, mass normalization $M_{200}$, transition
radius $r_{5/2}$, and interloper fraction $P_I$. \label{fig:conf}}
\end{center}
\end{figure*}

Figures \ref{fig:results} and \ref{fig:conf} show the results.  Aside
from difference in the value of $\gamma$, the high- and low- velocity
dispersion systems provide very similar fits: the galaxy orbits are
consistent with isotropic ($\tilde{\beta} = 0$), and both ensembles
have $n\approx2$ and transition radius $r_{5/2}$ much greater than
$\rtwo$. In other words, the matter distributions for both samples are
consistent with a singular isothermal sphere within the limits of the
survey.

\subsection{Discussion}

Neither the cluster nor the group data supports a total matter
distribution with a constant-density core ($n=0$). This result is in
keeping with recent studies of more distant clusters in the X-ray
\citep{Lewis03,Arabadjis02,Allen02} as well as weak lensing maps
\citep{Gavazzi03,Clowe02,Clowe00}. Optical observations of cluster
velocity data have also yielded similar results. For example,
\cite{Biviano03} uses the Jeans equation and isotropic velocity
dispersion profiles to rule out constant-density cores in the Two
Degree Field Galaxy Redshift Survey \citep{Colless01}. 

Our results suggest that within a projected region $R \lesssim
2\rtwo$, the matter distribution in both low- and high-$\sigma$
systems of galaxies is close to a single power law with $n=2$. The
density declines rapidly only outside $2\rtwo$. This result is
consistent with N-body simulations. For example, in the original paper
describing the simulations of NFW, the density profiles of dark matter
halos are consistent with a single power law of slope 2 between $0.1
\rtwo$ and $2 \rtwo$ (see their Figure 3). The only discrepancy, or
flattening, occurs at radii smaller than $0.1 \rtwo$, where our
datasets have 56 (low-$\sigma$) and 131 (high-$\sigma$) members,
perhaps insufficient to provide a robust constraint.

A major limitation of our method is the fixed form of the galaxy
distribution function $f(\ene,L^2)$. By modeling the DF as a power law
in energy and angular momentum, we neglect all other possible forms
for the DF, some of which may indeed be consistent with an $n=0$
model. On the other hand, the $\chi^2$ tests we perform reassure us
that the data are at least statistically consistent with cuspy matter
distributions.

The work of \cite{van00}, whose methods we adapt and extend, involves
a similar analysis, and it is instructive to compare the two different
approaches. Instead of fixing the form of the DF, they fix the form of
the surface number density $\Sigma(R)$, and calculate the DF using an
inversion of the integral in equation (\ref{eq:dfint}). As a result,
their DF is more general than ours, $f \propto g(\ene) L^{-2 \beta}$,
with $g$ being the result of the inversion. Their method has the
advantage of generality, but the disadvantage that it requires a
two-stage fit: first $\Sigma(R)$ must be fit and fixed, and then
$\beta$ and $n$ measured separately, without an indication of how
changes in the $\Sigma(R)$ fit would affect the resulting
constraints. Our method sacrifices generality by setting $g(\ene) =
\ene^{\alpha-1/2}$, but gains the advantage that all but one of the
parameters are fit simultaneously ($\gamma$ is varied discretely). As
a result, the correlation among the parameters is easy to understand
via marginalized confidence contours (Figure \ref{fig:conf}).

Another difference between our work and \cite{van00} is the sampling
of the parameter space. We constrain the continuous five-dimensional
region $(n,\beta,M_0,r_0,P_I)$, whereas \cite{van00} sample discrete
points within that space. They find that an NFW profile with $n=1$
matches the CNOC data, but they do not consider $n=2$ models; we find
that our clusters and groups have steeper matter distributions, with
the best fit model closer to $n=2$.

The fixed functional form of our DF can also explain the large
transition radii. Careful examination of equation (\ref{eq:nu}) shows
that increasing $\gamma$, the slope of the galaxy density as $r
\rightarrow \infty$, also steepens the slope of the galaxy density
$n_g$ near $r = 0$. It is still the case that $n_g = 2 \beta$ at $r =
0$, but the derivative of the slope $d n_g / d r$ becomes larger and
larger as $\gamma$ is increased. Our data have a steep galaxy density
slope near $r = 0$, and there are only two ways to fit this:
increasing $\beta$ (and thus generating more radially anisotropic
orbits) or increasing $\gamma$ (and thus steepening the overall galaxy
energy distribution and hence the galaxy density). This is where the
velocity distribution enters. As Figure \ref{fig:results} shows, the
grand total velocity histogram is not strongly peaked at $\tilde{v}_z
= 0$ (in fact, it is consistent with a Gaussian
distribution). Dramatically increasing $\beta$ would yield a velocity
distribution that is much more sharply peaked than the data. Thus the
only choice left the model is to increase $\gamma$ to larger values,
$\gtrsim 8$. However, the slope of the galaxy density is not as steep
as 8 at the limits of the survey.  Hence, a large transition radius
$r_0$ is required to fit the outer regions of the galaxy number
density profile. At the same time, the velocity data allow the matter
profile to be nearly a power law within the region constrained by the
data. Note that the upper limits on $\tilde{r}_{5/2}$ are possible
because if $\tilde{r}_{5/2}$ becomes too large, then the galaxy
density profile begins to resemble a single power law as well, and the
position data do not favor this limit (Figure \ref{fig:results}).

Another shortcoming is that our method constrains the total matter
density, rather than the dark matter density alone. While the dark
component is thought to dominate the mass, it does not do so
overwhelmingly---in typical clusters, $\sim 25\%$ of the mass is in
baryons, chiefly in the form of the X-ray emitting medium. In the
innermost regions of clusters with a dominant central galaxy, it is
actually the stellar mass density and not the dark matter density that
dominates the total distribution \citep{Koopmans03}. Lensing and
stellar dynamical models which take into account the starlight
separately from the dark component show conflicting results---while
the central regions of some clusters exhibit flat $n=0$ \citep{Sand02}
dark matter densities, others show evidence for steep $n \sim 2$ dark
matter profiles \citep{Davis03}.

It would be of great value to constrain the inner density slope $n$
independently of the slope at infinity. Unfortunately, because the
data are sparse, we cannot derive true constraints on the outer galaxy
density slope $\gamma$, and measuring the outer total matter density
would be even more difficult. With $10^5$ redshifts a true,
simultaneous constraint on the outer and inner slopes would be
possible; such large data sets could allow a direct measurement of the
phase space distribution function \citep{Merritt93b}, without
requiring us to parameterize the shape of the distribution function, as
in equation (\ref{eq:df}). Data sets that are only factors of two
larger are unlikely to make that difference. Also, leveraging
independent constraints on the inner and outer slopes by extending the
survey beyond $\rtwo$ is unlikely to provide better constraints. In
these outer regions, most of the galaxies are experiencing spherical
infall, which makes equilibrium models clearly unapplicable, and
caustic modeling techniques preferable
\citep{vanHaarlem93,Diaferio99,Geller99,Rines00}. However, one can
envision a technique that combines equilibrium dynamics at small
scales with the study of the infall caustics outside the virialized
regions of the cluster. In this way one could obtain a total mass
profile over several decades in $r$.

The most promising method of increasing the accuracy of this analysis
is to extend the known membership of groups and clusters to dwarf
galaxies with absolute magnitudes as faint as $M_R = -11$. A typical
group with velocity dispersion $\sigma \approx 300$ km s\m\ is likely
to have $\approx 60$ faint dwarf members within 1 Mpc
\citep{Trentham02}. A large spectroscopic survey of well-selected
galaxies brighter than $m_R = 22$ in our group sample would likely
yield triple the current membership count and allow for much more
robust constraints.

\section{Conclusion}
\label{sec:conclude}

We compare the distribution of matter in rich and poor systems of
galaxies, basing our analysis on a deep catalog of galaxies in
groups. The catalog consists of a statistically complete sample of
2419 newly measured redshifts in the fields of \ngroup\ systems, as
well as 979 redshifts from the literature for 8 nearby rich clusters
of galaxies. Most of the groups have X-ray emission detected in the
RASSCALS survey \citep{Mahdavi00}.

We construct a low-$\sigma$ ensemble group and a high-$\sigma$
ensemble cluster from these data. The final dynamical samples have 893
members in 33 groups, and 945 members in 15 clusters, with an average
velocity dispersion of 330 km s\m\ and 800 km s\m\, respectively.

Our study is the first to derive confidence volumes in the parameter
space defined by the matter density slope $n$, the interloper fraction
$P_I$, the mass normalization $M_{200}$, the velocity anisotropy
$\beta$, and the characteristic radius $r_0$ for clusters and groups
of galaxies. By modeling the phase space density of the galaxies
directly, we place much more stringent bounds on the mass and orbital
structure of these systems than would be possible through the virial
theorem or the Jeans equations. However, our results are limited by
the fact that we force the galaxy distribution function (DF) to be a
power law in energy and angular momentum, neglecting all other forms.

We find that neither the clusters nor the groups are consistent with a
matter distribution that has a flat, constant-density core ($n=0$).
Instead, within $R < 2\rtwo$, both the low-$\sigma$ and the
high-$\sigma$ systems are consistent with a single power law with
slope $n \approx 2$.  The models require both the matter and the
galaxy distribution to decline steeply outside 2 $\rtwo$. It is
conceivable that choices of DF different from ours could yield
substantially different results; however, the $\chi^2$ test cannot
reject the possibility that our model is correct. Our results are
consistent with cold CDM simulations that predict centrally divergent
matter distributions.

We thank the anonymous referee for detailed and insightful comments
that significantly improved the paper. We thank Perry Berlind, Michael
Calkins, and Susan Tokarz for observing and reducing most of the data
via the FAST pipeline. We have made use of the NASA/IPAC Extragalactic
Database (NED) which is operated by the Jet Propulsion Laboratory,
California Institute of Technology, under contract with the National
Aeronautics and Space Administration. This research was supported by
the Smithsonian Institution and by NASA through a Chandra Postdoctoral
Fellowship Award issued by the Chandra X-ray Observatory Center, which
is operated by the Smithsonian Astrophysical Observatory for and on
behalf of NASA under contract NAS 8-39073.


\singlespace
\renewcommand{\arraystretch}{0.9}

\begin{deluxetable}{lccrrr@{$\pm$}lr@{$\pm$}lr@{$\pm$}lc}
\tablewidth{0in}
\tabletypesize{\small}
\tableheadfrac{0.35}
\tablecaption{Newly Observed Systems \label{tbl:groups}}
\tablehead{\colhead{Name} & \colhead{$\alpha_{2000}$} & \colhead{$\delta_{2000}$} & $N_\mr{obs}$
& $N_\mr{memb}$ & \mc{$ c \bar{z}$ } & \mc{$ \sigma_z $} &
\mc{$\log{L_X}$}  & Other ID}
\startdata
SRGb062\tna & 00:18:25.2 & +30:04:13.8  &  84 &   45 &  6818 &  54 &   379 &  43& 42.66	& 0.14  \\
SRGb063     & 00:21:38.4 & +22:24:20.5  &  75 &   45 &  5771 &  48 &   336 &  36& 42.56	& 0.18 \\
SRGb090\tnb & 01:07:02.2 & +32:19:21.7  & 142 &  104 &  5150 &  47 &   508 &  34& 42.81	& 0.08 & NGC 383 Group\\
SRGb102\tnb & 01:25:28.8 & +01:45:17.0  & 109 &   35 &  5433 &  66 &   409 &  41& 42.56	& 0.12 & NGC 533 Group\\
SRGb119\tna & 01:56:21.6 & +05:37:04.7  &  54 &   29 &  5484 &  64 &   372 &  42& 42.30	& 0.20 & NGC 741 Group\\
SRGb145     & 02:31:48.0 & +01:16:27.2  &  55 &   37 &  6714 &  61 &   387 &  43& 42.34	& 0.31 \\
SRGb149     & 02:38:43.8 & +02:01:11.4  &  68 &   44 &  6670 &  54 &   376 &  38& \mc{$<41.98$} \\
SRGb155\tnb & 02:52:48.7 & -01:17:02.5  &  66 &   26 &  7259 & 123 &   650 &  93& 42.69 	& 0.20 \\
SRGb158     & 02:55:51.0 & +09:18:48.4  &  33 &   22 &  7810 &  61 &   301 &  36& \mc{$<41.98$} \\
NRGb004\tna & 08:38:11.5 & +25:07:00.1  &  37 &   20 &  8478 &  91 &   440 & 144& 42.937	& 0.14 \\
NRGb007\tna & 08:49:31.9 & +36:35:40.1  &  28 &   11 &  7625 &  94 &   322 &  81& \mc{$<41.979$} \\
NRGb025\tna & 09:13:28.3 & +30:06:13.7  &  35 &   31 &  6735 &  81 &   458 &  65& \mc{$<41.86$} \\
NRGs027    & 09:16:06.2 & +17:36:08.3  &  59 &   36 &  8615 &  63 &   387 &  54& 42.62	& 0.18 \\
NRGb032\tna& 09:19:48.0 & +33:45:32.8  &  67 &   45 &  6834 &  64 &   459 &  63& 42.69	& 0.14 & Abell 779\\
NRGs038\tnb & 09:23:35.0 & +22:19:58.4  &  65 &   56 &  9218 & 104 &   784 &  60& 42.47	& 0.24 \\
NRGb043\tna& 09:28:09.9 & +30:03:35.0  &  29 &   25 &  7884 &  47 &   257 &  32& \mc{$<42.07$} \\
NRGb045\tna& 09:33:27.1 & +34:03:02.9  &  26 &    9 &  8211 &  36 &   110 &  31& 42.11	& 0.42 \\
NRGb057\tna & 09:44:28.1 & +36:08:55.7  &  24 &   13 &  6766 &  45 &   168 &  25& \mc{$<41.76$} \\
SS2b144     & 09:49:59.9 & -05:02:48.3  &  35 &   21 &  6303 &  43 &   202 &  22& \mc{$<41.88$} \\
NRGs076 \tnb& 10:06:41.8 & +14:25:49.8  &  60 &   40 &  9273 & 150 &   929 & 145& 42.34	& 0.30 \\
NRGb078     & 10:13:53.5 & +38:44:24.4  &  87 &   39 &  6770 &  51 &   335 &  35& 42.16	& 0.52 \\
NRGs110\tnb & 11:00:50.9 & +10:33:17.3  &  60 &   38 & 10579 &  76 &   493 &  62& 42.75	& 0.20 & Abell 1142\\
NRGs117\tnb & 11:10:31.4 & +28:43:39.4  &  86 &   80 &  9814 &  73 &   667 &  47& 42.98	& 0.13 & Abell 1185 \\
NRGs127\tna & 11:21:30.5 & +34:13:21.5  &  12 &   10 & 10485 &  64 &   206 &  36& \mc{$<42.13$} \\
SS2b164    & 11:22:43.0 & -07:44:54.2  &  88 &   47 &  7107 &  51 &   361 &  46& 42.33	& 0.27 \\
NRGs156\tna& 11:46:16.3 & +33:11:03.1  &  35 &   28 &  9640 &  58 &   320 &  37& \mc{$<41.97$} \\
NRGb177     & 12:04:17.8 & +20:15:18.0  & 113 &   74 &  6995 &  48 &   416 &  35& 42.52	& 0.16 & NGC 4065 Group\\
NRGb181\tna & 12:07:07.4 & +31:19:46.5  &  17 &   13 &  7053 & 118 &   426 &  70& \mc{$<41.78$} \\
NRGb184     & 12:08:01.0 & +25:15:13.7  & 113 &   52 &  6726 &  48 &   376 &  40& 42.30	& 0.24 \\
NRGs241\tnb & 13:20:16.6 & +33:08:13.2  &  59 &   40 & 10980 &  78 &   503 &  45& 42.92	& 0.12 \\
NRGb244\tna & 13:24:10.8 & +13:58:47.6  &  33 &   19 &  6947 &  56 &   268 &  33& 42.63	& 0.18 & NGC 5129 Group\\
NRGb247     & 13:29:31.2 & +11:47:19.0  &  58 &   39 &  6850 &  61 &   395 &  43& 42.67	& 0.18 & NGC 5171 Group\\
NRGb251\tna & 13:34:25.7 & +34:40:54.8  &  48 &   26 &  7345 &  50 &   265 &  30& 42.44	& 0.17 \\
SS2b239     & 13:49:10.8 & -07:18:11.7  &  69 &   24 &  7329 &  59 &   311 &  29& 42.38	& 0.26 & HCG 67\\
NRGb302     & 14:28:33.1 & +11:22:07.7  &  53 &   31 &  7892 &  58 &   324 &  32& 42.05	& 0.59 \\
NRGs317\tna & 14:47:09.8 & +13:42:23.4  &  39 &   19 &  8898 &  72 &   338 &  38& 42.51	& 0.30 \\
SRGb009\tna & 22:14:48.0 & +13:50:17.5  &  46 &   35 &  7775 &  57 &   349 &  33& 42.44	& 0.26 \\
SRGb013\tnb & 22:50:00.7 & +11:40:15.6  &  55 &   30 &  7689 & 101 &   578 &  76& 42.45	& 0.21 \\
SRGb016     & 22:58:14.2 & +25:56:13.0  &  58 &   43 &  7394 &  46 &   327 &  33& \mc{$<41.84$} \\
SRGb037     & 23:28:46.6 & +03:30:49.1  &  85 &   20 &  5221 &  86 &   413 &  59& 41.91	& 0.44 \\
SS2b312     & 23:47:24.0 & -02:19:08.4  &  59 &   27 &  6773 &  47 &   274 &  42& 42.41	& 0.20 & HCG 97\\
\enddata
\tablecomments{$N_\mr{obs}$ is the number of galaxies who redshifts were measured in this work; $N_\mr{memb}$ is the
number we identify as group members; $\bar{z}$ is the mean group redshift, and $\sigma_z$ is the velocity dispersion
in km s\m, with errors from standard bootstrap analysis.
 $L_X$ is the 0.1-2.4 keV luminosity in $\hsq$ erg s\m as calculated
in \cite{Mahdavi00}.}
\tablenotetext{a}{The data for these groups appeared originally in \cite{Mahdavi99}; however, our algorithm for
determining group membership differs from the one used in that work, and hence the total number
of members varies slightly. Also, $N_\mr{obs}$ in our table differs from $N_\mr{Total}$ in \cite{Mahdavi99},
because we list the actual number of galaxies observed, rather than the number in the field
 brighter than a given magnitude.
}
\tablenotetext{b}{Because of their large velocity dispersions, these systems were grouped with the rich clusters in Table \ref{tbl:clusterstwo} for our analysis.}
\end{deluxetable}

\begin{deluxetable}{lrrrrrrrrrc}
\tablecaption{New Galaxy Positions and Velocities \label{tbl:gals}\tna}
\tablehead{\colhead{Name} & \mct{$\alpha_{2000}$} &
\mct{$\delta_{2000}$} & \colhead{$c z$} & \colhead{$\epsilon_{c z}$} &
\colhead{$d$} & \colhead{Member}}
\tablewidth{0in}
\startdata
SRGb062.001 &00 &11 &57.0 &+29 &29 &08 &28038  &35  &1.808  &N \\
SRGb062.002 &00 &12 &11.8 &+29 &19 &09 & 7728  &21  &1.838  &N \\
SRGb062.003 &00 &12 &17.5 &+29 &52 &17 & 6921  &20  &1.595  &N \\
SRGb062.004 &00 &12 &28.4 &+29 &32 &38 & 6862  &15  &1.656  &N \\
SRGb062.005 &00 &12 &38.3 &+30 &06 &08 & 6791  &23  &1.487  &M \\
SRGb062.006 &00 &12 &45.0 &+29 &22 &15 &10419  &15  &1.683  &N \\
SRGb062.007 &00 &13 &12.7 &+31 &08 &43 &14465  &28  &1.846  &N \\
SRGb062.008 &00 &13 &45.1 &+30 &11 &40 & 7109  &28  &1.209  &M \\
\enddata
\tablecomments{$c z$ is the galaxy redshift times the
speed of light in km s\m, $\epsilon_{c z}$ is
the measurement error, and $d$ is the distance from the
X-ray center in Mpc assuming
$H_0$ = 100 km s\m\ Mpc\m. Members are indicated by an ``M''
in the Member column; nonmembers are indicated by an ``N.''}
\tablenotetext{a}{This table is for illustrative purposes only.
The complete data appear in the electronic version of the
Journal.}
\end{deluxetable}

\begin{deluxetable}{cr@{:}r@{:}rr@{:}r@{:}rrr@{$\pm$}lr@{$\pm$}lrrc}
\tablewidth{0in}
\tabletypesize{\small}
\tablecaption{Cluster Sample \label{tbl:clusterstwo}}
\tablehead{\colhead{Name} & \mct{$\alpha_{2000}$} & \mct{$\delta_{2000}$} 
& $N_\mr{memb}$ & \mc{$ c \bar{z}$ } & \mc{$ \sigma_z $} & $\log{L_X}$ & References }
\startdata
Abell 262     & 01&52&50.4 &  36&08&46 & 103 &   4920 & 55 & 560 & 41  & 43.14 &  0.58  & G97,N01 \\
Abell 3158    & 03&42&39.6 & -53&37&50 & 118 &  17740 & 91 & 964 & 56  & 44.12 &  1.70  & K01,G96 \\  
Abell 496     & 04&33&37.1 & -13&14&46 & 227 &   9920 & 53 & 729 & 37  & 43.95 &  1.17  & D00,M99 \\
Abell 1644    & 12&57&14.8 & -17&21&13 &  85 &  14210 &100 & 947 & 73  & 43.94 &  \nodata  & G97 \\
Abell 1795    & 13&49&00.5 &  26&35&07 &  84 &  18880 & 95 & 879 & 77  & 44.44 &  1.76  & G97 \\
Abell 1809    & 13&53&18.9 &  05&09&15 &  61 &  23750 &100 & 747 & 67  & 43.60 &  \nodata  & G97 \\
Abell 3571    & 13&47&28.9 & -32&51&57 &  90 &  11700 &101 & 969 & 69  & 44.26 &  1.72  & G96 \\
Abell 2029    & 15&10&58.7 &  05&45&42 &  87 &  23140 &140 &1235 & 78  & 44.58 &  2.20  & O95,S94 \\
\enddata
\tablecomments{$N_\mr{memb}$ is the
number we identify as cluster members; $\bar{z}$ is the mean cluster redshift, 
and $\sigma_z$ is the velocity dispersion of our derived membership
in km s\m, with errors from standard bootstrap analysis.
$L_X$ is the total 0.1-2.4 keV luminosity in $\hsq$ erg s\m from \cite{Ebeling96},
the typical error is 20\%.
The references are to X-ray and optical studies of the clusters that
suggest they could be dynamically relaxed.
C97: \cite{Cirimele97}; D00: \cite{Durret00}; 
G96: \cite{Girardi96}; G97: \cite{Girardi97}; K01: \cite{Kolo01}; 
O95: \cite{Oegerle95};
N01: \cite{Neill01}; S94: \cite{Slezak94}.}
\end{deluxetable}

\newcommand{\mcf}[1]{\multicolumn{4}{c}{#1}}

\begin{deluxetable}{cccccccccccccccc}
\tablewidth{0in}
\tabletypesize{\small}
\tableheadfrac{0.35}
\tablecaption{Simulations of the Measurement Technique \label{tbl:sims}}
\tablehead{& \mc{$\tilde{\beta}$} && \mc{$n$} 
&& \mc{$\tilde{r}_{5/2}$} && \mc{$\tilde{M}_{200}$} && \mc{$P_I$} & Fit \\
\cline{2-3} \cline{5-6} \cline{8-9} \cline{11-12} \cline{14-15}
\colhead{ID} & \colhead{In} & \colhead{Out}  && \colhead{In} & \colhead{Out}  && \colhead{In} & \colhead{Out}  && \colhead{In} & \colhead{Out}  && \colhead{In} & \colhead{Out} & Quality }
\startdata
A & 0.0 & -0.3\, -- \, 0.3  && 1.0 & 0.6--1.3 && 1.0 & 0.8--1.2 && 0.3 & 0.27--0.35 && 0.05 & 0.04--0.07 & 0.96 \\
B & 0.5 & -0.2\, -- \, 0.7  && 2.0 & 1.9--2.3 && 1.5 & 1.2--1.8 &&  3  & 1.3--5.0   && 0.1  & 0.08--0.15 & 0.75 \\
C &-0.5 & -1.0\, -- \,-0.3  && 0.0 & 0.0--1.3 && 2.0 & 1.8--2.2 &&  1  & 0.6--2.0   && 0.01 & 0.0--0.03  & 0.20 \\
D & 0.0 & -0.5\, -- \, 0.2  && 0.0 & 0.0--0.7 && 1.0 & 0.7--1.0 && 0.4 & 0.3--0.4   && 0.03 & 0.01--0.04 & 0.34
\enddata
\tablecomments{Simulations of 893 galaxies carried out with $\gamma=8$. ``In'' refers to the input parameter
set; ``Out'' shows the recovered 1D 95\% marginalized confidence interval for each parameter.}
\end{deluxetable}

\begin{deluxetable}{lcc}
\tablewidth{0in}
\tabletypesize{\small}
\tableheadfrac{0.35}
\tablecaption{Summary of Results \label{tbl:props}}
\tablehead{\colhead{} & \colhead{Low-$\sigma$} & \colhead{High-$\sigma$}}
\startdata
\mct{\emph{General Statistics}} \\ 
No. of systems  &   33  &  15       \\
Completeness limit & $1.8 \rtwo$ & $0.8 \rtwo$  \\
No. of members within $0.8\rtwo$  &  521  & 945     \\
No. of members within $1.8\rtwo$  &  893  & \nodata     \\
Range in $\sigma_z$ (km/s)  & 110-470  & 470-1240\\ 
Gaussian fit to $\tilde{v}_z$ & $\chi^2/\nu = 22/27$ & $\chi^2/\nu = 32/28$ \\
\\
KS Test in $R$\tna  & \mc{$d = 0.106, p = 10^{-3}$} \\
KS Test in $\tilde{v}_z$  & \mc{$d = 0.028, p = 0.87$} \\
\\
\mct{\emph{Likelihood of Constant Density Core ($n=0$)}} \\ 
Fit Quality, $\beta \le 0$  & $q = 0.003$ & $q < 10^{-5}$ \\
Fit Quality, $\beta$ free  & $q = 0.02$ & $q = 0.003$ \\
\\
\mct{\emph{Likelihood of Free $n$}}\\ 
Minimum $\gamma$    & 8 & 12 \\
Fit Quality        & $q = 0.169$ & $q = 0.124$ \\
$n$         & \grouprange & \clusterrange \\
$\tilde{\beta}$      & -0.37 \, -- \, 0.42 & 0.02\, --\, 0.67  \\
$r_{5/2}/\rtwo$ & 3--12 & 5.8--26 \\
$G M_{200}/\sigma^2 r_0$ & 2.2--2.7  &  2.2--3.0 \\
$P_I$       &  0.01--0.05    & 0.01--0.02 \\
\enddata
\tablenotetext{a}{Conducted only for galaxies with $R < 0.8 \rtwo$.}
\end{deluxetable}

\renewcommand{\arraystretch}{1}

\clearpage

\appendix
\section{The Distribution Function Integral}
\label{sec:appendix}

In the statistical interpretation of the distribution function (DF),
the probability of observing a galaxy with a line-of-sight peculiar
velocity $v_z$ at a projected distance $R$ from the cluster center in
a survey of $N_0$ galaxies within a limiting radius $\rlim$ is
\citep{Merritt93,van00}
\begin{equation}
p^\prime(R,v_z) = \frac{2 \pi R}{N_0} 
\int \int \int f(\ene,L^2) d v_\phi \, d v_R \, d z,
\end{equation}
where ($v_R$,$v_\phi$,$v_z$) are the components of the galaxy's
peculiar motion in cylindrical coordinates, $\ene \equiv \Psi(r) -
v^2/2$, $L$ is the angular momentum, and $f$ is the distribution
function (DF). The probability distribution is normalized such that
\begin{equation}
\int_{0}^{R_\mr{lim}} \int_{-\infty}^{\infty} p^\prime(R,v_z|\mb{a})
\, dv_z \, dR = \int_{0}^{R_\mr{lim}} \frac{2 \pi R}{N_0} \Sigma(R) \,
dR = 1,
\label{eq:apnorm}
\end{equation}
where $\Sigma(R)$ is the surface number density of galaxies.  We use
a DF of the form
\begin{equation}
f(\ene,L^2) = f_0 \ene^{\alpha-1/2} L^{-2 \beta}
\end{equation}
where $f_0$ is fixed by the normalization requirement in equation
(\ref{eq:apnorm}). The limits of the integration are determined
as follows: (1) a galaxy bound to the cluster must always have
$v_R^2+v_\phi^2+v_z^2 \le 2 \Psi$, and (2) a galaxy with peculiar
velocity $v_z$ has an associated maximum radius $r_\mr{max}$ such that
$v_z^2 = 2 \Psi(r_\mr{max})$, beyond which it is unbound. With these
conditions, and switching to the coordinate system $v_R = w
\sin{\eta}$, $v_\phi = w \cos{\eta}$, we have
\begin{eqnarray}
p^{\prime}(R,v_z) & = & \frac{2 \pi R f_0}{N_0}
\int_R^{r_\mr{max}} \frac{2 r dr}{\sqrt{r^2 - R^2}}
\int_0^{2 \pi} d \eta \times \\
& & \int_0^{\sqrt{2 \Psi - v_z^2}} 
\left[ \Psi(r) - w^2/2 - v_z^2/2 \right]^{\alpha - 1/2}
L^{-2 \beta}
w dw, \nonumber \\
L^2 & = & r^2 w^2 \cos^2{\eta} + r^2 
 \left( w \sin{\eta} \cos{\theta} - v_z \sin{\theta} \right)^2, \\
\sin{\theta} & \equiv & \frac{R}{r}.
\end{eqnarray}
For $\beta > 0$, the above expression has an integrable singularity
at $L = 0$, or $(\eta = \pi/2,w = v_z \tan{\theta})$. Changing variables
from $(\eta,w)$ to $(Q,L^2)$ is tempting, but the integration
boundaries in those coordinates are not necessarily analytic. It is
best to work in the $(\eta,w)$ plane and take advantage of the fact
that, for Gaussian quadrature, the innermost integrand will never be
evaluated at the singular point $\eta = \pi/2$, and so will always be
finite at $w = v_z \tan{\theta}$. We find that making the change of
variables appropriate for an inverse square-root singularity in $w$
($\omega = w^2$) \citep{NR} yields accurate results for the innermost
integral. As for the middle, $\eta$ integral, it may be evaluated
using Gaussian quadrature \citep{NR} with a constant weighting
function for $\beta < 1/2$. For $\beta \ge 1/2$, Gaussian
quadrature, with $(\pi/2 - \eta)^{1-2 \beta}$ as the weighting
function, is best. 

The galaxy number density $\nu(r)$ corresponding to the chosen DF is
given by an integral over all velocities; this integral is discussed in
detail by \cite{Cuddeford} and \cite{Merritt85a}.
\begin{eqnarray}
\nu(r) & = & \int_{v^2 \le 2 \psi} f(\ene,L^2) d^3 \mb{v} 
\\ & = & \frac{2 \pi f_0}{r} \int_0^\Psi \ene^{\alpha-1/2} d \ene 
\int_0^{2 r^2 (\Psi-\ene) }
\frac{L^{-2 \beta} d L^2 }{\sqrt{2 r^2 (\Psi - \ene) - L^2}}
\\ & = & 
2^{3/2-\beta} \pi^{3/2} f_0 \frac{\Gamma(\alpha+1/2) 
\Gamma(1-\beta)}{\Gamma(\alpha-\beta+2)} r^{-2
  \beta} \Psi(r)^{\alpha-\beta+1}.
\end{eqnarray}

\end{document}